\documentclass[aps,prl,twocolumn,10pt,superscriptaddress,nofootinbib,nobibnotes,longbibliography]{revtex4-1}

\usepackage{amssymb}
\usepackage{graphicx}
\usepackage{amsmath}
\usepackage{mathrsfs}
\usepackage[colorlinks=true,linkcolor=blue,citecolor=blue,urlcolor=blue]{hyperref}
\usepackage{subfigure}
\usepackage{multirow}
\usepackage{setspace}
\usepackage{verbatim}
\usepackage{float}
\usepackage{color}
\usepackage{ulem}
\usepackage[utf8]{inputenc}
\usepackage[table,xcdraw]{xcolor}
\usepackage{makecell}
\usepackage{url}

\newcommand{\black}{\textcolor{black}}

\begin{document}

\title{Scalar-induced gravitational waves with non-Gaussianity up to all orders}

\author{Xiang-Xi Zeng}
\email{zengxiangxi@itp.ac.cn}
\affiliation{Institute of Theoretical Physics, Chinese Academy of Sciences (CAS), Beijing 100190, China}
\affiliation{School of Physical Sciences, University of Chinese Academy of Sciences (UCAS), Beijing 100049, China}

\author{Zhuan Ning}
\email{ningzhuan17@mails.ucas.ac.cn}
\affiliation{School of Fundamental Physics and Mathematical Sciences, Hangzhou Institute for Advanced Study (HIAS), University of Chinese Academy of Sciences (UCAS), Hangzhou 310024, China}
\affiliation{Institute of Theoretical Physics, Chinese Academy of Sciences (CAS), Beijing 100190, China}
\affiliation{University of Chinese Academy of Sciences (UCAS), Beijing 100049, China}

\author{Rong-Gen Cai}
\email{caironggen@nbu.edu.cn}
\affiliation{Institute of Fundamental Physics and Quantum Technology, \& School of Physical Science and Technology, Ningbo University, Ningbo, 315211, China}

\author{Shao-Jiang Wang}
\email{schwang@itp.ac.cn}
\affiliation{Institute of Theoretical Physics, Chinese Academy of Sciences (CAS), Beijing 100190, China}
\affiliation{Asia Pacific Center for Theoretical Physics (APCTP), Pohang 37673, Korea}

\begin{abstract}
Scalar-induced gravitational waves (SIGWs) are ubiquitous in many early-Universe processes accompanied by non-Gaussianity; while Gaussian perturbation can generate significant SIGWs, computations of SIGWs can be significantly affected and enhanced if the scalar perturbations have some degree of non-Gaussianity; hence, precise calculations of these kinds of SIGWs involve a full understanding of non-Gaussianity. In this Letter, we propose to use the lattice simulations to directly calculate the energy density spectra of SIGWs with non-Gaussianity up to all orders. Our proposal has been first verified to match the existing semi-analytical results with non-Gaussianity, and then applied to more general cases, including high-order primordial non-Gaussianities, the logarithmic dependence in curvature perturbations, the curvaton model, and the ultra slow-roll model. We find that even a modest non-Gaussianity can significantly alter ultraviolet behaviors in SIGW spectra, necessitating special cautions in future detections as well as mutual constraints on/from primordial black holes.
\end{abstract}
\maketitle

\textit{\textbf{Introduction.}---} 
Scalar-induced gravitational waves (SIGWs)~\cite{Baumann:2007zm, Ananda:2006af} are a prominent source of gravitational radiation and can encode rich information about early-universe physics (see Ref.~\cite{Domenech:2021ztg} for a review), including the associated production of primordial black holes (PBHs)~\cite{Saito:2008jc, Saito:2009jt}. Producing an observable SIGW requires an enhancement of curvature perturbations but on those scales that do not affect cosmic microwave background (CMB) observables; such enhancement can arise in scenarios like ultra-slow-roll inflation~\cite{Ivanov:1994pa, Leach:2000ea, Alabidi:2012ex, Ballesteros:2017fsr, Ballesteros:2020qam, Germani:2017bcs, Bhaumik:2019tvl, Ragavendra:2020sop, Garcia-Bellido:2017mdw} or the curvaton model~\cite{Lyth:2001nq, Enqvist:2001zp, Moroi:2001ct, Lyth:2002my, Sasaki:2006kq, Pi:2021dft} (see Ref.~\cite{LISACosmologyWorkingGroup:2025vdz} for a review). These mechanisms typically induce \black{nonlinearities}, also known as non-Gaussianity.

A standard approach to investigate this non-Gaussianity is to expand the exact curvature perturbation $\zeta$ in terms of a Gaussian field $\zeta_g$~\cite{Gangui:1993tt, Komatsu:2001rj},
\begin{align}\label{eq: expand}
    \zeta = \zeta_g + F_{\mathrm{NL}}\zeta_g^2 + G_{\mathrm{NL}}\zeta_g^3 + H_{\mathrm{NL}}\zeta_g^4 + \cdots,
\end{align}
where $F_{\mathrm{NL}}, G_{\mathrm{NL}}, H_{\mathrm{NL}}, \ldots$ parametrize the amplitude of non-Gaussianity. By transforming the perturbations to momentum space, a semi-analytic formalism has been established~\cite{Kohri:2018awv, Espinosa:2018eve}, allowing an order-by-order calculation of the non-Gaussianity, but at the price of progressively higher-dimensional integrals that are not feasible in practice.

Numerous studies have investigated high-order non-Gaussian effects on the SIGW energy spectrum using this formalism~\cite{Cai:2018dig, Unal:2018yaa, Adshead:2021hnm, Li:2023xtl, Yuan:2023ofl, Li:2025met, Perna:2024ehx}, and their cosmological implications have also been widely studied in literature~\cite{Zhou:2025djn, Luo:2025lgr, Picard:2024ekd, Cai:2024dya, Han:2025wlo, Pearce:2025ywc, Cecchini:2025oks, Domenech:2025bvr, Zeng:2025ecx, Yu:2025jgx, Chianese:2025mll, Ghaleb:2025xqn, Peng:2024ktq, Ragavendra:2021qdu, Ragavendra:2025svk}. Particularly, even the first-order non-Gaussian parameter $F_{\mathrm{NL}}$ can significantly impact the PBH abundance~\cite{Ozsoy:2023ryl, Byrnes:2012yx, Young:2013oia, Biagetti:2021eep, Atal:2018neu, Taoso:2021uvl, Young:2022phe, Ferrante:2022mui, Gow:2022jfb, Papanikolaou:2024kjb, He:2024luf}, highlighting the importance of precise calculation of SIGW spectra to constrain the PBH abundance. Due to the difficulty of high-dimensional integrals, they are limited to providing results only up to the $G_{\mathrm{NL}}$ order\footnote{Although the expansion in Ref.~\cite{Perna:2024ehx} includes terms up to the fourth order, their calculation does not account for all loop effects.} and it is impractical to obtain fully \black{non-Gaussian} results. On the other hand, recent work~\cite{Iovino:2024sgs} has shown that truncating Eq.~\eqref{eq: expand} at finite order can not capture the right amplitude of the SIGW energy spectrum.

Motivated by these limitations and by previous numerical simulations of SIGWs in an early matter-dominated era~\cite{Fernandez:2023ddy}, we propose here a complementary approach: performing lattice simulations of the fully \black{non-Gaussian} curvature perturbations in coordinate space, computing the resulting tensor perturbations directly, and then transforming the outcomes back to momentum space to obtain the \black{final} SIGW energy spectrum.

\textit{\textbf{Scalar-induced gravitational waves.}---}
In the conformal Newtonian gauge, the perturbed metric reads
\begin{align}
    \mathrm{d}s^2 = a^2(\eta)\left\{ -(1+2\Phi)\mathrm{d}\eta^2 +  \left[(1-2\Phi)\delta_{ij} + \frac{1}{2}h_{ij} \right]\mathrm{d}x^i\mathrm{d}x^j  \right\},
\end{align}
where $\Phi$ denotes the scalar perturbation (we neglect anisotropic stress so $\Phi = \Psi$), and $h_{ij}$ is the transverse-traceless (TT) tensor perturbation. Expanding the Einstein equations to the second order yields
\begin{align}
    h_{ij}^{\prime\prime} + 2\mathcal{H}h_{ij}^{\prime} - \nabla^2h_{ij} = - 4 S_{ij}^{\mathrm{TT}},
\end{align}
with primes indicating derivatives with respect to the conformal time $\eta$ and $\mathcal{H} = a^{\prime}/a$ the conformal Hubble parameter. Here, $S_{ij}^{\mathrm{TT}}$ is the TT projection of the quadratic source~\cite{Baumann:2007zm},
\begin{align}
    S_{ij} =& 4\Phi\partial_i\partial_j\Phi + 2\partial_i\Phi\partial_j\Phi \nonumber\\
    &- \frac{4}{3(1+\omega)\mathcal{H}^2}\partial_i(\Phi^{\prime} + \mathcal{H}\Phi)\partial_j(\Phi^{\prime} + \mathcal{H}\Phi),
\end{align}
where $\omega\equiv p/\rho$ is the background equation-of-state (EoS) parameter. For the adiabatic case, the perturbation $\Phi$ obeys
\begin{align}
    \Phi^{\prime\prime} + 3\mathcal{H}(1+c_s^2)\Phi^{\prime} + 
    (2\mathcal{H}^{\prime} + (1+3c_s^2)\mathcal{H}^2 - c_s^2\nabla^2)\Phi = 0,  
\end{align}
with the sound speed satisfying $c_s^2 = \omega$. In what follows, we solve these equations by lattice simulations in coordinate space.

\textit{\textbf{Simulation setup.}---}
For adiabatic perturbation with enhanced power on scales larger than the initial Hubble horizon $\mathcal{H}^{-1}$, the initial scalar perturbation $\Phi_i$ is related to the curvature perturbation $\zeta_i$ by
\begin{align}
    \Phi_i = \frac{3+3\omega}{5+3\omega}\zeta_i,
\end{align}
with the super-horizon freezing condition $\Phi^{\prime}_i \simeq 0$~\cite{Mukhanov:2005sc} (the subscript `$i$' denotes the initial time). We set the tensor initial conditions in the usual way, $(h_{ij})_{i} = (h_{ij}^{\prime})_i = 0$. The initial curvature perturbation $\zeta_i$ should be a generic nonlinear functional $F$ of a Gaussian random field $\zeta_g$,
\begin{align}
    \zeta = F[\zeta_g].
\end{align}

The cosmological Gaussian random field~\cite{Mukhanov:2005sc} is most naturally specified in momentum space. We adopt the convention
\begin{align}
    \zeta_g(\vec{x}) = \int \frac{\mathrm{d}^3x}{(2\pi)^3} \zeta_g(\vec{k})
    e^{i\vec{k}\cdot\vec{x}},
\end{align}
and assume the Fourier modes are complex Gaussian variables with probability density
\begin{align}
    \mathscr{P}[\zeta_g(\vec{k})] = \frac{1}{\pi\sigma_{k}^2} \mathrm{exp}\left[-\frac{|\zeta_g(\vec{k})|^2}{\sigma_{k}^2}\right].
\end{align}
The two-point correlation is then
\begin{align}
    \langle \zeta_g(\vec{k})\zeta_g(\vec{q}) \rangle = \sigma_k^2 \delta^3(\vec{k} + \vec{q}),
\end{align}
so that $\sigma_k^2 = (2\pi)^3P_{\zeta_g}(k)$, where $P_{\zeta_g}(k)$ is the dimensional power spectrum of $\zeta_g$. 

We perform simulations in a comoving cubic box with periodic boundary conditions, computing spatial derivatives with a Fourier Pseudo-Spectral (FPS) method and advancing in time with a fourth-order Runge-Kutta integrator. The tensor evolution and TT projection follow the procedures of Refs.~\cite{Garcia-Bellido:2007fiu, Dufaux:2007pt, Figueroa:2011ye}. The GW energy spectrum before redshifted to today is defined as usual by~\footnote{We have verified in our simulation that $\langle h_{ij,k}h_{ij,k} \rangle$ = $\langle h_{ij}^{\prime}h_{ij}^{\prime} \rangle$.}
\begin{align}
    \Omega_{\mathrm{GW}} = \frac{\rho_{\mathrm{GW}}}{\rho_c(\eta)} = \frac{1}{48\mathcal{H}^2}\langle h_{ij}^{\prime}h_{ij}^{\prime} \rangle,
\end{align}
where angle brackets denote a volume average and $\rho_c(\eta)$ is the critical energy density at time $\eta$.

\begin{figure}
    \centering
    \includegraphics[width=0.5\textwidth]{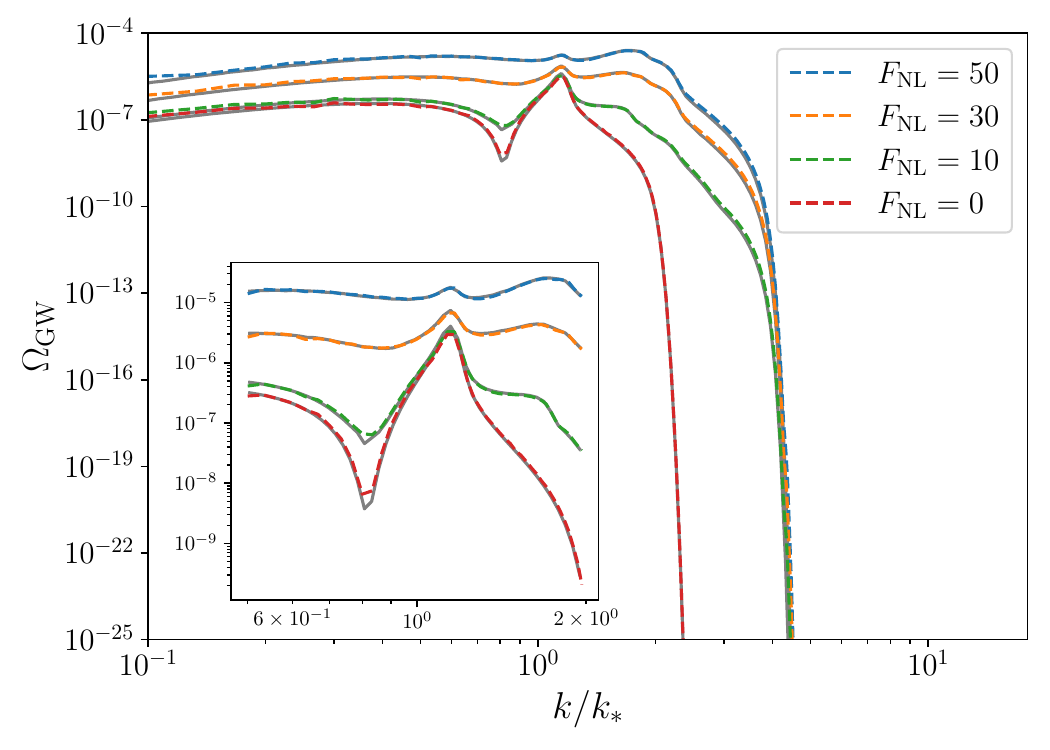}
    \caption{Energy spectra of SIGW for different non-Gaussian parameters $F_{\mathrm{NL}} = 0, 10, 30, 50$. Dashed curves are simulation results with a lattice number $N^3 = 256^3$; solid gray curves show semi-analytical results computed with \textbf{vegas}~\cite{Lepage:2020tgj}.}
    \label{fig:fnl}
\end{figure}

To validate the code before tackling the fully \black{non-Gaussian} task, we reproduce semi-analytical results~\cite{Adshead:2021hnm} using a Gaussian power spectrum with a bump,
\begin{align}\label{eq:gaussPS}
    \mathcal{P}_{\zeta_g}(k) = A\frac{(k/k_*)^3}{\sqrt{2\pi}e}\mathrm{exp}\left[ -\frac{(k/k_* - 1)^2}{2e^2} \right].
\end{align}
Concretely we choose $A=10^{-3}, e=1/30, \omega=1/3$ and consider the first-order non-Gaussianity $\zeta = \zeta_g  + F_{\mathrm{NL}}(\zeta_g^2 - \langle \zeta_g^2\rangle )$. As shown in Fig~\ref{fig:fnl}, the numerical results (dashed curves) closely match the semi-analytical predictions (solid gray curves) in this simple case. Convergence tests \black{with larger lattice numbers $N^3=512^3, 1024^3$ adopting leapfrog and finite-difference methods} are attached in the \textit{Supplemental Material}, \black{where agreements among different methods also serve as consistency checks for the FPS method we adopted in the main text.}

\textit{\textbf{Simulation results.}---}
We perform all simulations assuming a radiation-dominated background ($\omega=1/3$); generalizations to other values of $\omega$ should be straightforward. We focus on four representative non-Gaussian prescriptions:  (1) the high-order non-Gaussianity; (2) the logarithmic relation of $\zeta$~\cite{Iovino:2024sgs, Pi:2022ysn}; (3) the curvaton model~\cite{Sasaki:2006kq}; and (4) the ultra slow-roll model with an upward step~\cite{Cai:2022erk}. Except for the case (1), we adopt the Gaussian bump power spectrum~\eqref{eq:gaussPS} with $A=10^{-2}, e=10^{-1}$ for typical illustrations. \black{More convergence tests for each case are attached in the \textit{Supplemental Material}.}

\begin{figure}
    \centering
    \includegraphics[width=0.5\textwidth]{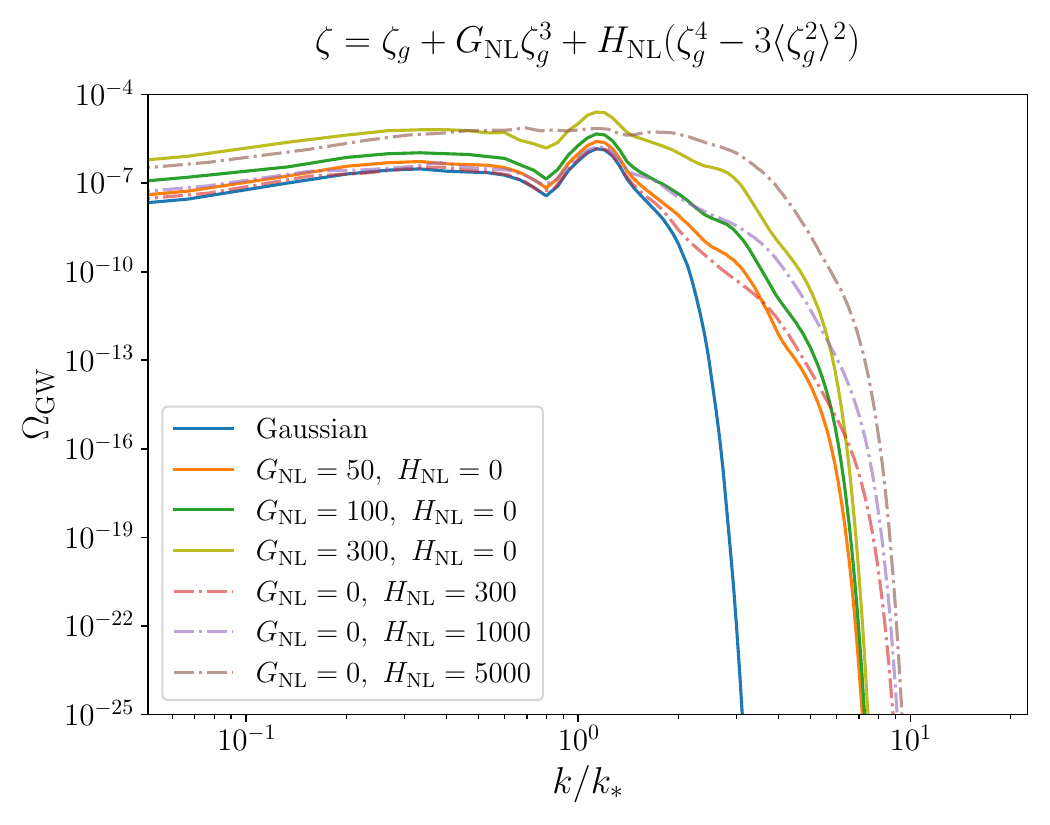}
    \caption{Energy spectra of SIGW for different higher-order non-Gaussian parameters $G_{\mathrm{NL}} = 50, 100, 300$ and $H_{\mathrm{NL}} = 300, 1000, 5000$. We use the Gaussian bump power spectrum~\eqref{eq:gaussPS} with $A=10^{-3}, e=1/10$ and the lattice number \black{$N^3=256^3$}.}
    \label{fig:hignoeder}
\end{figure}

(1) High-order non-Gaussianity. As a first example, we include cubic and quartic contributions,
\begin{align}
    \zeta = \zeta_g + G_{\mathrm{NL}}\zeta_g^3 + H_{\mathrm{NL}}(\zeta_g^4 - 3\langle  \zeta_g^2\rangle^2 ).
\end{align}
which, in semi-analytical treatment, requires evaluating up to 12- and 16-point correlators. As shown in Fig.~\ref{fig:hignoeder}, higher order terms tend to add power at higher frequencies---a consequence of the conservation of momentum~\cite{Zeng:2024ovg, Cai:2019amo}. Therefore, it is expected that the nonlinear spectra exhibit markedly different ultraviolet behavior compared with low-order results, which has also been found in the next three cases.

\begin{figure}
    \centering
    \includegraphics[width=0.5\textwidth]{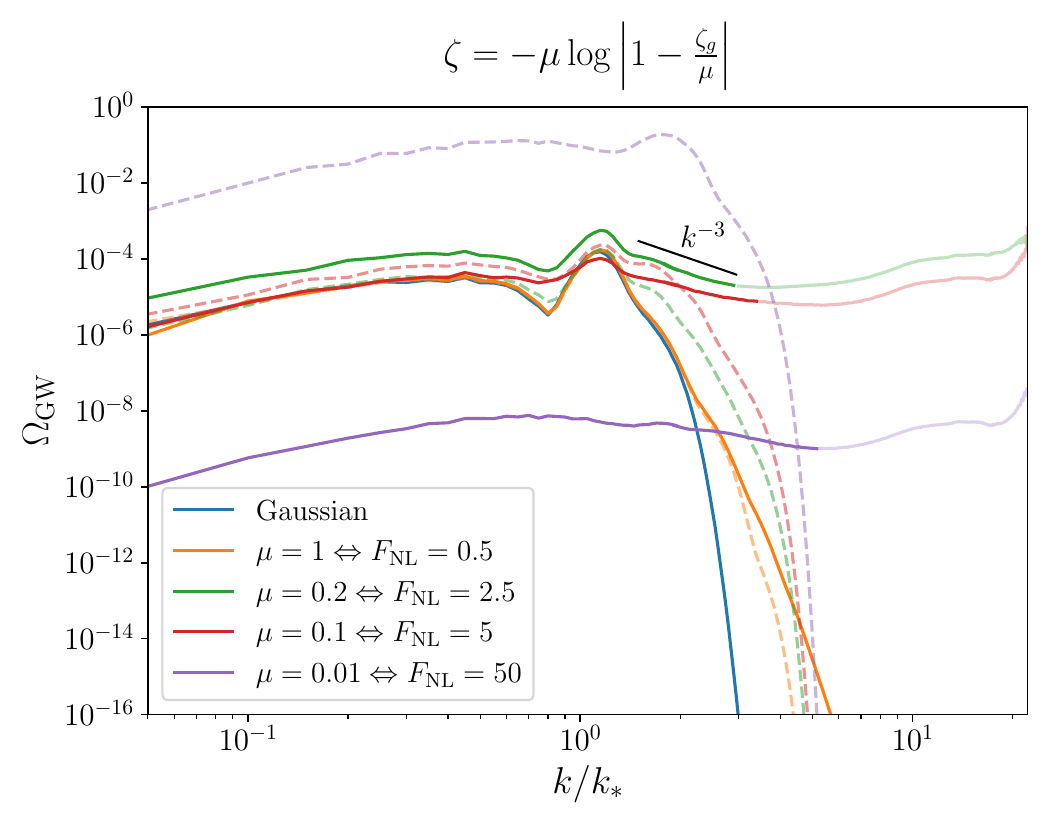}
    \caption{Energy spectra of SIGW with different $\mu$ defined in \eqref{eq:logzeta}. The solid and dashed curves denote the nonlinear results and the perturbative results up to $F_{\mathrm{NL}}$, respectively. The lattice number in the simulation is $N^3=256^3$. The shallow solid curve describes the frequency band where numerical simulations fail, as detailed in the \textit{Supplemental Material}.}
    \label{fig:logrelation}
\end{figure}

(2) Logarithmic relation of $\zeta$. Following the notation of Ref.~\cite{Iovino:2024sgs}, we consider a logarithmic dependence in $\zeta_g$,
\begin{align}\label{eq:logzeta}
    \zeta = - \mu \log \left| 1 - \frac{\zeta_g}{\mu} \right|,
\end{align}
where $\mu$ is the parameter controlling the amplitude of non-Gaussianity, which depends on the specific model. To leading order, one finds
\begin{align}
    F_{\mathrm{NL}} = \frac{1}{2\mu}.
\end{align}
In Fig.~\ref{fig:logrelation}, solid and dashed curves represent nonlinear results and perturbative results truncated at $F_{\mathrm{NL}}$, respectively. It can be seen that even modest non-Gaussianity changes the ultraviolet tail of the SIGW spectra into an approximate power-law form, and varying $\mu$ can enhance or suppress the amplitude of $\Omega_{\mathrm{GW}}$ relative to the perturbative prediction, with orders of magnitude difference in GW amplitude due to a large $F_\mathrm{NL}$ between solid and dashed purple curves. Furthermore, sufficiently large non-Gaussianity can even shift the spectral peak.

\begin{figure}
    \centering
    \includegraphics[width=0.5\textwidth]{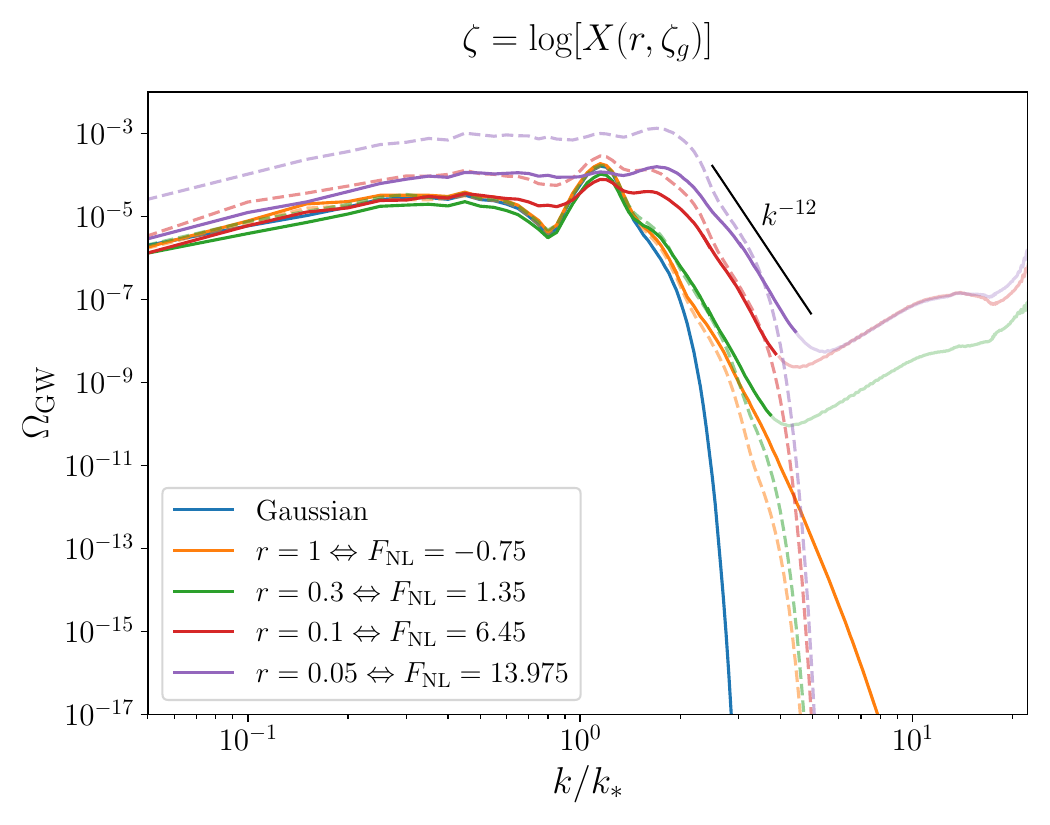}
    \caption{Energy spectra of SIGW for different $r$ defined in \eqref{eq:curvaton1}. The solid and dashed curves denote the nonlinear results and the perturbative results up to $F_{\mathrm{NL}}$, respectively. The lattice number in the simulation is $N^3=256^3$. The shallow solid curve describes the frequency of numerical simulation failure. }
    \label{fig:curvaton}
\end{figure}

(3) Curvaton model. An analytical functional form of the curvature perturbation for the curvaton mechanism is derived in Ref.~\cite{Sasaki:2006kq} as
\begin{align}\label{eq:curvaton1}
    \zeta = \log[X(r, \zeta_g)],
\end{align}
where $r$ denotes the energy fraction of the curvaton to the total energy density at the time of curvaton decay. The explicit expressions of $X$ and $r$ are given in the \textit{Supplementary Material}. After expanding with $\zeta_g$, one finds
\begin{align}
    F_{\mathrm{NL}} = \frac{3}{4r} - 1 - \frac{r}{2}.
\end{align}
Fig.~\ref{fig:curvaton} compares nonlinear spectra (solid) with perturbative spectra truncated at $F_{\mathrm{NL}}$ (dashed). Similar to the logarithmic case, the nonlinear spectra exhibit an approximate power-law behavior at high frequencies, and the amplitude and peak frequency can be significantly altered by non-Gaussianity. These effects are not captured reliably by the perturbative expansion.

\begin{figure}
    \centering
    \includegraphics[width=0.5\textwidth]{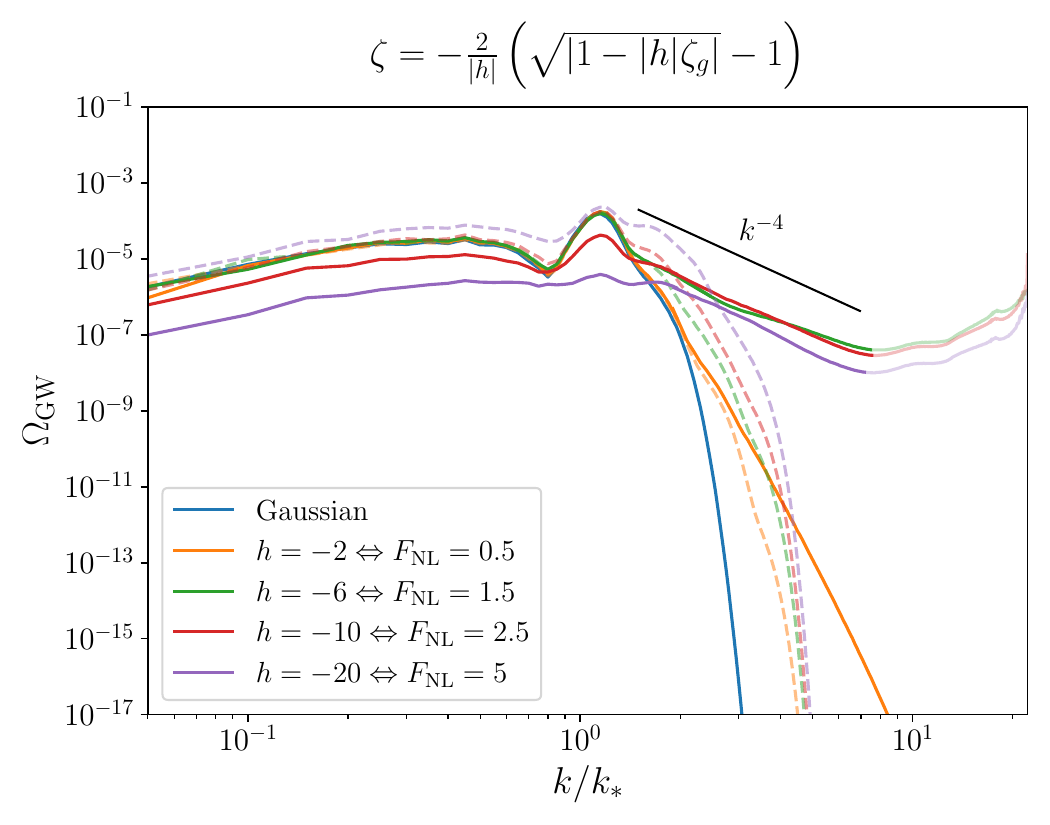}
    \caption{Energy spectra of SIGW for different $h$ defined in \eqref{eq:step}. The solid and dashed curves denote the nonlinear results and the perturbative results up to $F_{\mathrm{NL}}$, respectively. The lattice number in the simulation is $N^3=256^3$. The shallow solid curve describes the frequency of numerical simulation failure. }
    \label{fig:step}
\end{figure}

(4) Ultra slow-roll model with an upward step. If the inflation potential contains a tiny upward step at the end of the non-attractor stage of the inflaton, the curvature perturbation behaves like~\cite{Cai:2022erk}
\begin{align}
    \zeta = -\frac{2}{|h|}\left(\sqrt{1 - |h|\zeta_g} - 1 \right),
\end{align}
where $h$ can be any negative value. This model naturally bounds the curvature perturbation by $\zeta \leq 2/|h|$ and can thus alleviate PBH overproduction~\cite{Wang:2024nmd}. Therefore, it is important to investigate its SIGW signal. Allowing for regions where $|h|\zeta_g > 1$ we implement the mapping as
\begin{align}\label{eq:step}
    \zeta = -\frac{2}{|h|}\left(\sqrt{\Big|1 - |h|\zeta_g\Big|} - 1 \right).
\end{align}
It is easy to find, up to the first order,
\begin{align}
    F_{\mathrm{NL}} = \frac{|h|}{4}.
\end{align}
In this model, although $F_{\mathrm{NL}}$ is typically small, $\zeta$ itself is highly nonlinear. We plot the energy spectra of SIGW in Fig.~\ref{fig:step} for different values of $|h|$. It is clear that increasing $|h|$ tends to reduce the GW amplitude in the nonlinear treatment, whereas a truncation at $F_{\mathrm{NL}}$ alone yields the opposite trend. This discrepancy highlights the importance of considering the full nonlinearity in SIGW calculations, as well as the constraints imposed by/on PBHs.

Finally, we compare the three nonlinear cases above with the same first-order amplitude $F_{\mathrm{NL}} = 5$ in Fig.~\ref{fig:compare}. The comparison highlights that different nonlinear prescriptions produce distinct signatures in $\Omega_{\mathrm{GW}}$, suggesting that precise measurements of SIGW spectra could discriminate among models of primordial non-Gaussianity.

\begin{figure}
    \centering
    \includegraphics[width=0.5\textwidth]{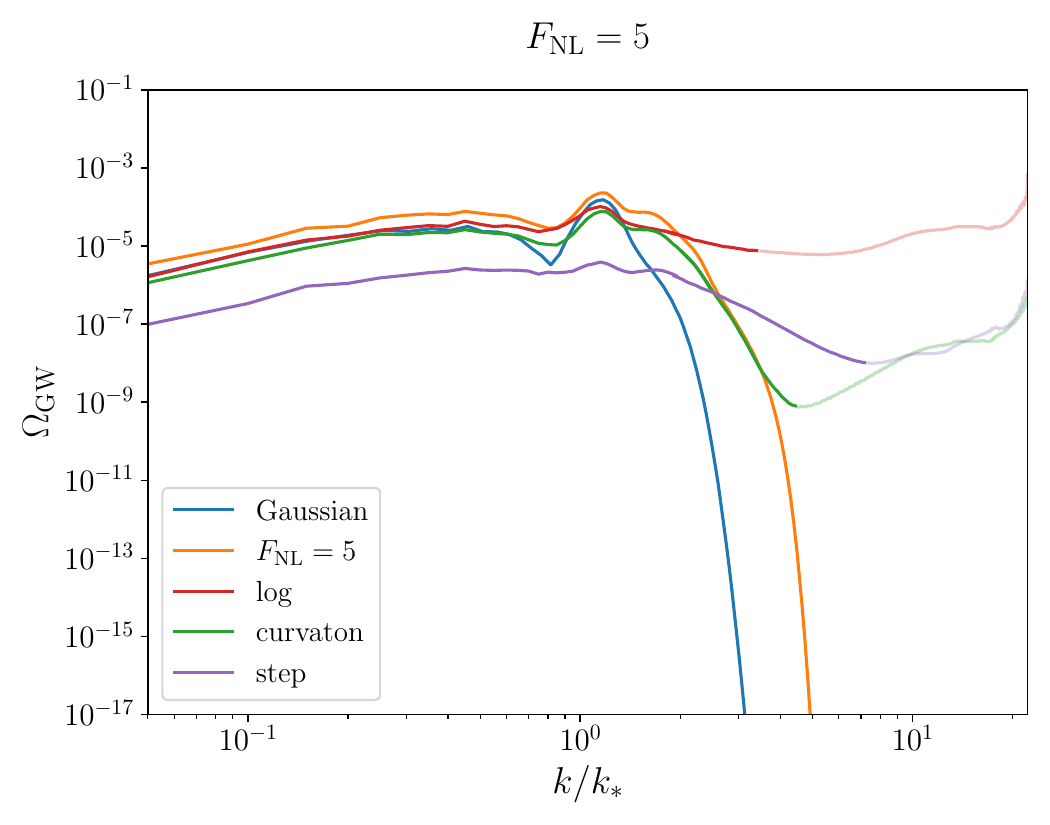}
    \caption{Energy spectra of SIGW for different models. The first-order amplitude has been set to $F_{\mathrm{NL}} = 5$ in all three models. The lattice number in the simulation is $N^3=256^3$. The shallow solid curve describes the frequency of numerical simulation failure. }
    \label{fig:compare}
\end{figure}

\textit{\textbf{Conclusions and discussions.}---}
Accurate prediction of the SIGW energy spectrum is crucial as the next generation of space-borne GW detectors like LISA~\cite{LISA:2017pwj}, Taiji~\cite{Hu:2017mde, Ruan:2018tsw}, and TianQin~\cite{TianQin:2015yph} are to be launched. Improved observational precision will not only enable discrimination between different early-universe scenarios but also help constrain the abundance of PBHs, for which the correct determination of the SIGW peak frequency is particularly important to estimate PBH masses.

In this \textit{Letter}, we propose to use lattice simulations to directly evolve the scalar perturbations to obtain their SIGW spectra with non-Gaussianity up to all orders. Applying this approach to several representative mappings between $\zeta$ and a Gaussian field $\zeta_g$, we find that even modest non-Gaussianity can substantially modify the ultraviolet behavior of the GW spectrum; larger nonlinearity can further change the amplitude and shift the spectral peak. On the other hand, a perturbative treatment truncated at low order can give misleading results. Our results also suggest far-reaching consequences for previous mutual constraints between SIGWs and PBHs.

The approach is flexible and can be improved or extended in several directions. First, by combining our pipeline with numerical-inflation techniques~\cite{Caravano:2024tlp, Caravano:2024moy, Caravano:2025diq} one can compute SIGW spectra directly from inflationary dynamics to improve accuracy. Although we illustrated results using a peak-like power spectrum, the method generalizes easily to broad spectra. It also accommodates non-local-type non-Gaussianity of the form
\begin{align}
    \zeta = F[\zeta_g, \nabla \zeta_g,  \nabla^2 \zeta_g, \cdots].
\end{align}
Therefore, how to deduce the nonlinear probability distribution function of $\zeta$ becomes essentially crucial for future investigations. Second, extension to isocurvature scenarios is likewise possible~\cite{Zeng:2025tno}, where it is hard to obtain the exact transfer function of the scalar perturbation $\Phi$~\cite{Domenech:2021and, Domenech:2020ssp, Kodama:1986ud, Domenech:2023jve}.
Finally, a caveat is that when the amplitude of curvature perturbation is extremely large, non-perturbative effects (higher-order perturbations) become important~\cite{Chang:2023aba} and the evolution of scalar perturbation may differ qualitatively from perturbative expectations~\cite{Inomata:2020cck, DeLuca:2023tun}. Addressing that regime likely requires full numerical-relativity simulations; we defer such an extension to future work.

\textit{\textbf{Data availability.}---}
\black{The Python/PyTorch code with the Fourier Pseudo-Spectral method is now available at \href{https://github.com/ZengXiang-Xi/SimuSIGW/tree/main}{\texttt{SimuSIGW}}, which is suitable for implementing numerical simulations with $N^3=256^3$ on a personal computer with GPU accelerations.}

\begin{acknowledgments}
\textit{\textbf{Acknowledgements.}---}
We thank Zi-Yan Yuwen for some useful discussions. 
This work is supported by the National Key Research and Development Program of China Grants No. 2021YFC2203004, No. 2021YFA0718304, and No. 2020YFC2201501, 
the National Natural Science Foundation of China Grants No. 12422502, No. 12547110, No. 12588101, No. 12235019, and No. 12447101,
and the China Manned Space Program Grant No. CMS-CSST-2025-A01.
\end{acknowledgments}

\bibliography{ref}


\onecolumngrid
\newpage
\appendix

\begin{center}
{\Large\textbf{Supplemental Material: Scalar-induced gravitational waves with non-Gaussianity up to all orders}}
\end{center}

\begin{center}
{
Xiang-Xi Zeng$^{1,2}$, Zhuan Ning,$^{3,1,4}$ Rong-Gen Cai$^{5}$ and Shao-Jiang Wang$^{1,6}$
}
\end{center}

\begin{center}
{\small \textit{$^1$Institute of Theoretical Physics, Chinese Academy of Sciences (CAS), Beijing 100190, China}}
\end{center}

\begin{center}
{\small \textit{$^2$School of Physical Sciences, University of Chinese Academy of Sciences (UCAS), Beijing 100049, China}}
\end{center}

\begin{center}
{\small \textit{$^3$School of Fundamental Physics and Mathematical Sciences, Hangzhou Institute for Advanced Study (HIAS), University of Chinese Academy of Sciences (UCAS), Hangzhou 310024, China}}
\end{center}

\begin{center}
{\small \textit{$^4$University of Chinese Academy of Sciences (UCAS), Beijing 100049, China}}
\end{center}

\begin{center}
{\small \textit{$^5$Institute of Fundamental Physics and Quantum Technology, \& School of Physical Science and Technology, Ningbo University, Ningbo, 315211, China}}
\end{center}

\begin{center}
{\small \textit{$^6$Asia Pacific Center for Theoretical Physics (APCTP), Pohang 37673, Korea}}
\end{center}

\section{Appendix A. Convergence tests \black{on the Gaussian case}}
\begin{figure}[h]
    \centering
    \includegraphics[width=0.48\linewidth]{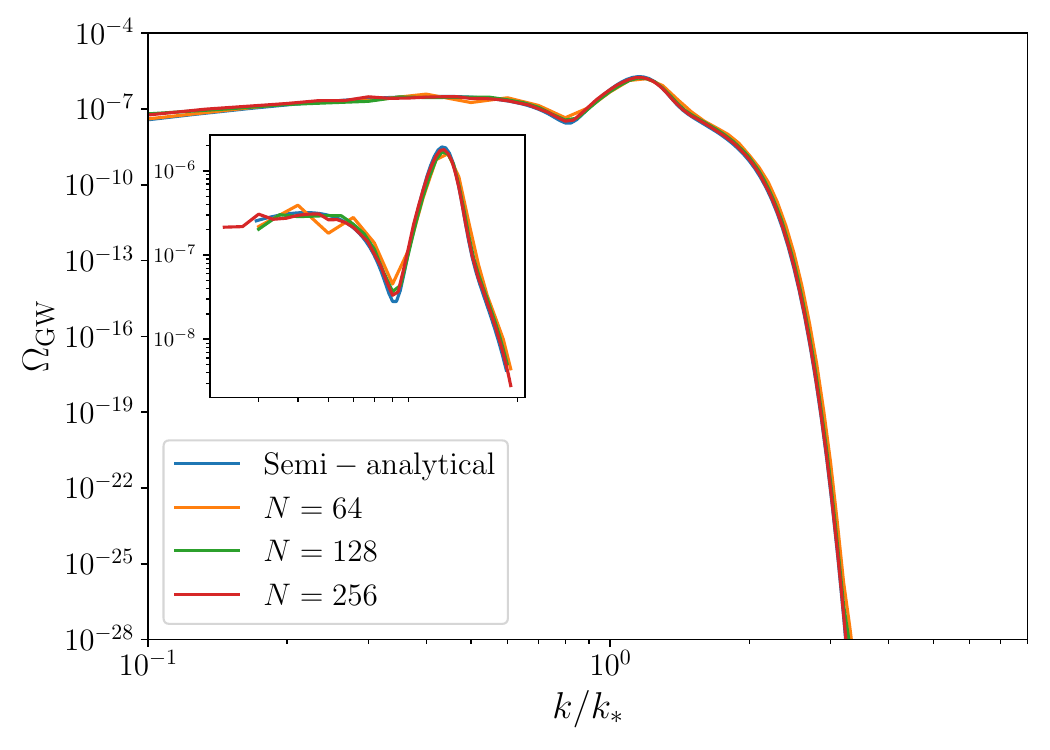}
    \hspace{0.1in}
    \includegraphics[width=0.48\linewidth]{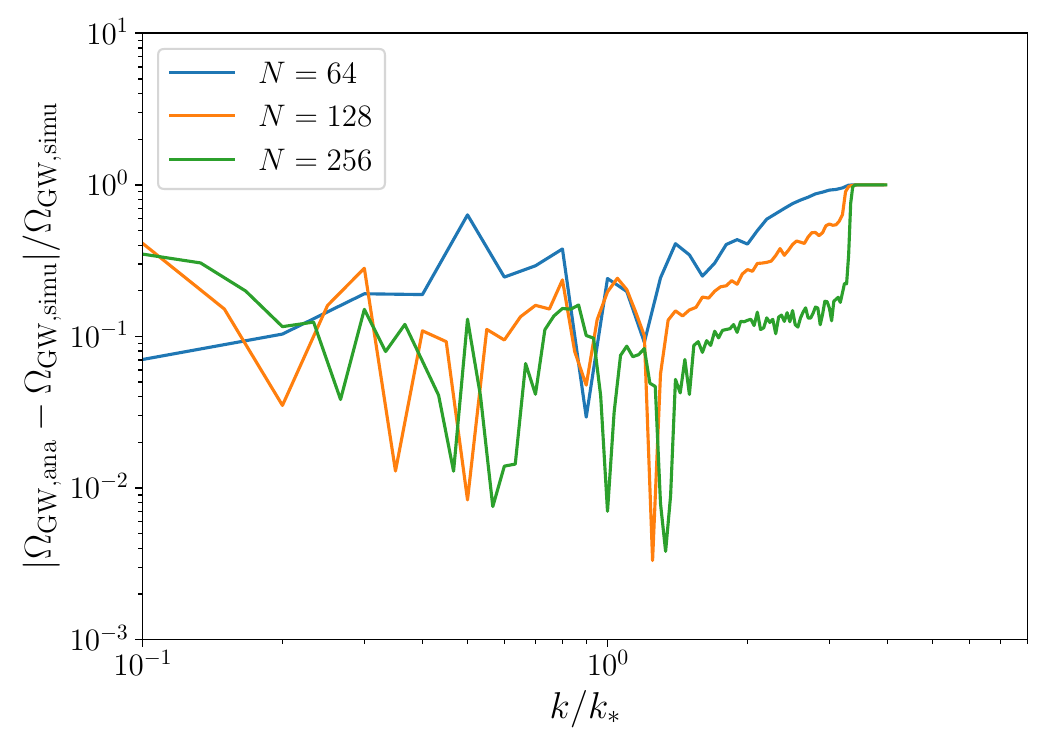}
    \caption{\textit{Left column}: Energy spectra of SIGW from simulations with different lattice numbers $N^3=64^3, 128^3, 256^3$ compared to the semi-analytical result. \textit{Right column}: The relative difference $|\Omega_{\mathrm{GW,semi-analytical}} - \Omega_{\mathrm{GW, simulation}}|/\Omega_{\mathrm{GW,simulation}}$ for the same lattice numbers.}
    \label{fig:testerror}
\end{figure}

We \black{first} use a Gaussian curvature perturbation as an example to test the convergence of our numerical simulations. The power spectrum is assumed to be
\begin{align}
    \mathcal{P}_{\zeta_g}(k) = A\frac{(k/k_*)^3}{\sqrt{2\pi}e}\mathrm{exp}\left[ -\frac{(k/k_* - 1)^2}{2e^2} \right].
\end{align}
with $A = 10^{-3}, e=0.1$. The left panel of Fig.~\ref{fig:testerror} shows the energy spectra of SIGW from simulations with different lattice numbers $N^3=64^3, 128^3, 256^3$ compared to the semi-analytical result. It can be seen that even a relatively coarse grid ($N^3 = 64^3$) reproduces the qualitative spectral shape; increasing the resolution to $128^3$ and $256^3$ improves quantitative agreement. In the right panel, we compare the semi-analytical result, which itself is just an approximated result, with respect to our numerical simulations (considered here as the exact results), $|\Omega_{\mathrm{GW,semi-analytical}} - \Omega_{\mathrm{GW, simulation}}|/\Omega_{\mathrm{GW, simulation}}$. One can see that the overall shape of the semi-analytical result is determined better than $10\%$ and even better than $1\%$ around the peak frequencies.

To further cross-validate our findings, we have applied the leapfrog method to improve the lattice number used by the Fourier Pseudo-Spectral method up to $N=512$. We have also developed a second, independent simulation pipeline based on the Finite Difference (FD) method (modified from the public code \texttt{CosmoLattice}~\cite{Figueroa:2020rrl, Figueroa:2021yhd, Ning:2025yvj}). This CPU-parallelized code allows us to run simulations with resolutions up to $N = 1024$. 

\begin{figure}[htbp]
    \centering
    \includegraphics[width=0.5\linewidth]{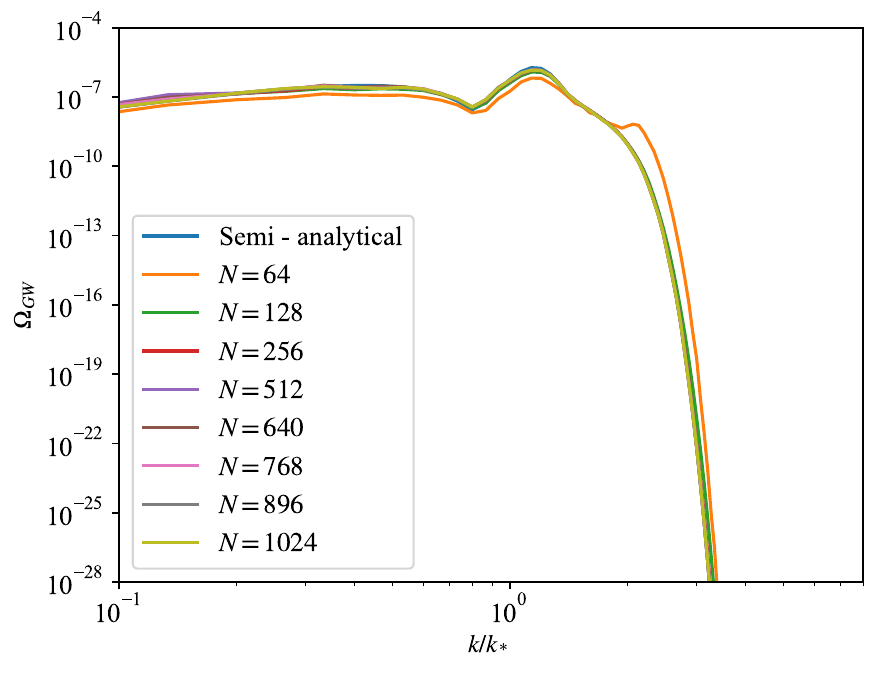}\\
    \includegraphics[width=0.5\linewidth]{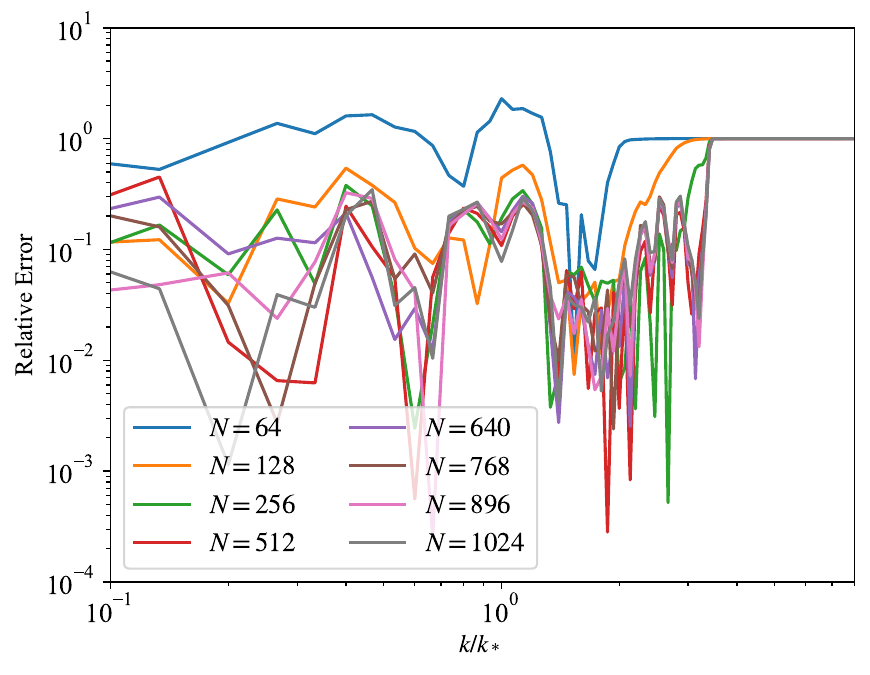}\\
    \includegraphics[width=0.5\linewidth]{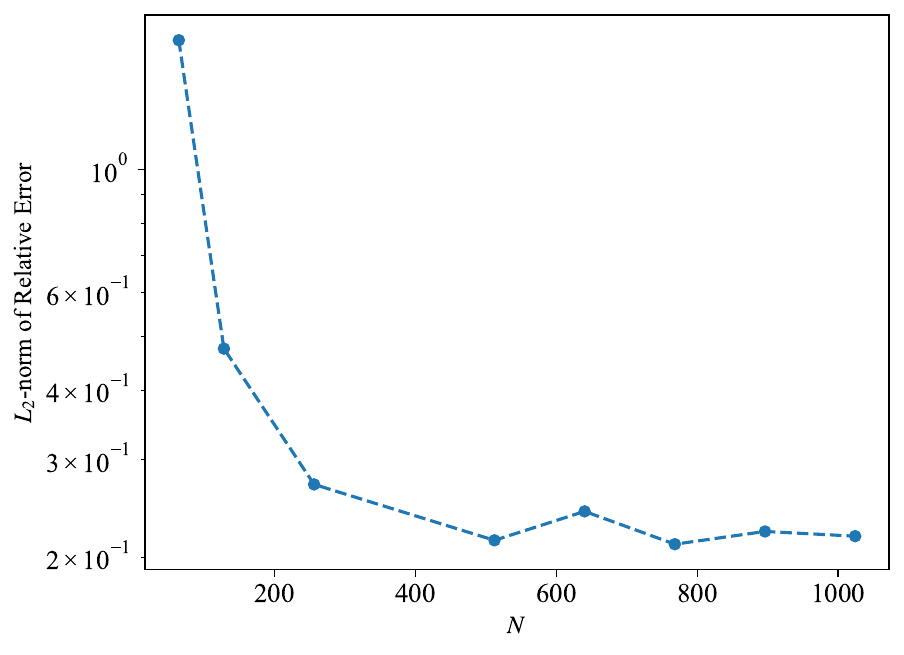}\\
    \caption{\black{Convergence tests for the benchmark Gaussian case using the Finite Difference (FD) code. Top: Energy spectra $\Omega_{\text{GW}}$ for different lattice resolutions ($N=64$ to $N=1024$) compared to the semi-analytical prediction (solid blue line). Middle: Relative error $|\Omega_{\text{GW, ana}} - \Omega_{\text{GW, simu}}|/\Omega_{\text{GW, simu}}$ as a function of frequency for each resolution. Bottom: Normalized $L_2$-norm of the relative error $\mathcal{E}_{L_2}(N)$ as a function of lattice resolution, showing clear convergence behavior.}}
    \label{fig:Gaussian_FD}
\end{figure}

We first use the benchmark Gaussian case to establish a convergence baseline using the high-resolution FD code. In the top panel of Fig.~\ref{fig:Gaussian_FD}, we plot the energy spectra $\Omega_{\text{GW}}$ obtained from simulations with lattice numbers ranging from $N=64$ to $N=1024$. The numerical results visually converge to the semi-analytical prediction (solid blue line) across the relevant frequency range. To quantify this agreement, the middle panel of Fig.~\ref{fig:Gaussian_FD} displays the relative error between the simulation and the semi-analytical solution, defined as $|\Omega_{\text{GW, ana}} - \Omega_{\text{GW, simu}}|/\Omega_{\text{GW, simu}}$. As the resolution increases, the relative error decreases significantly across the spectrum. To provide a rigorous scalar metric for convergence, we computed the normalized $L_2$-norm of the relative error over the Fourier modes, defined as:
\begin{equation}
    \mathcal{E}_{L_2}(N) = \sqrt{\frac{\sum_k [\Omega_{\text{GW, ana}}(k) - \Omega_{\text{GW, simu}}(k)]^2}{\sum_k [\Omega_{\text{GW, simu}}(k)]^2}}
\end{equation}
As shown in the bottom panel of Fig.~\ref{fig:Gaussian_FD}, this global error metric drops rapidly as $N$ increases and exhibits a clear convergence behavior, stabilizing around $N=512$. 

\section{Appendix B. Convergence tests on non-Gaussian cases}\label{app:B}

\begin{figure}[htbp!]
    \centering
    \includegraphics[width=0.32\linewidth]{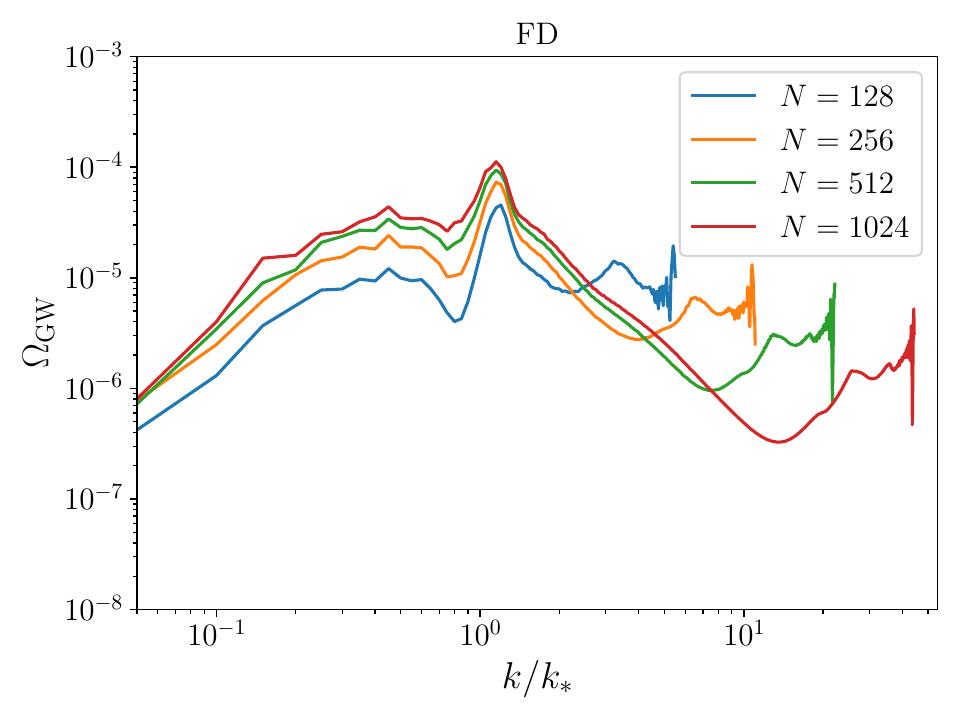}
    \includegraphics[width=0.32\linewidth]{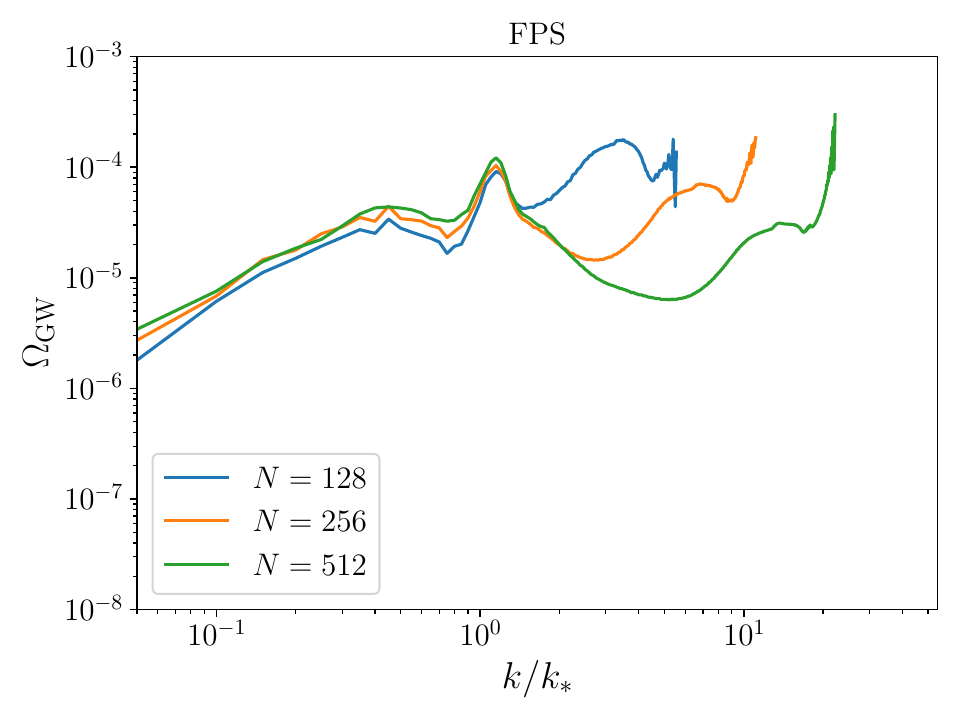}
    \includegraphics[width=0.32\linewidth]{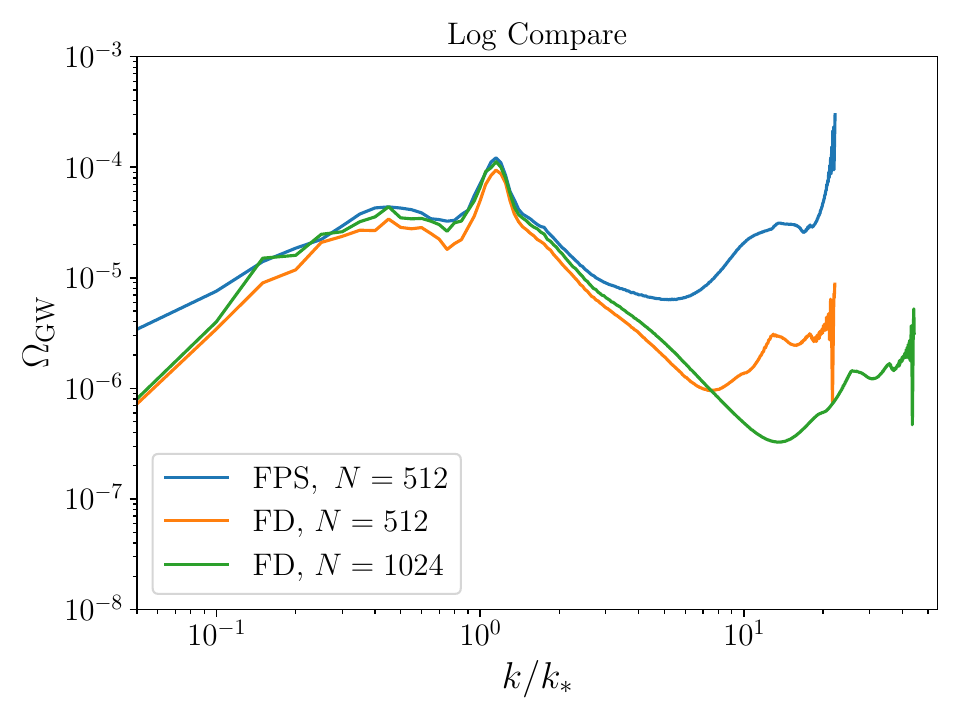}\\
    \includegraphics[width=0.32\linewidth]{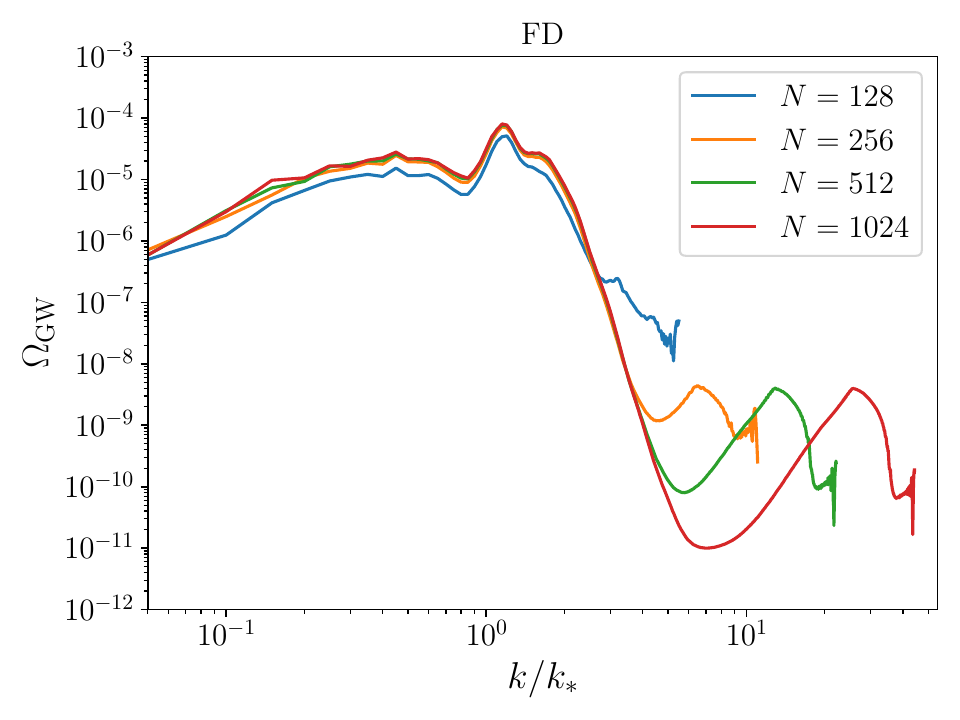}
    \includegraphics[width=0.32\linewidth]{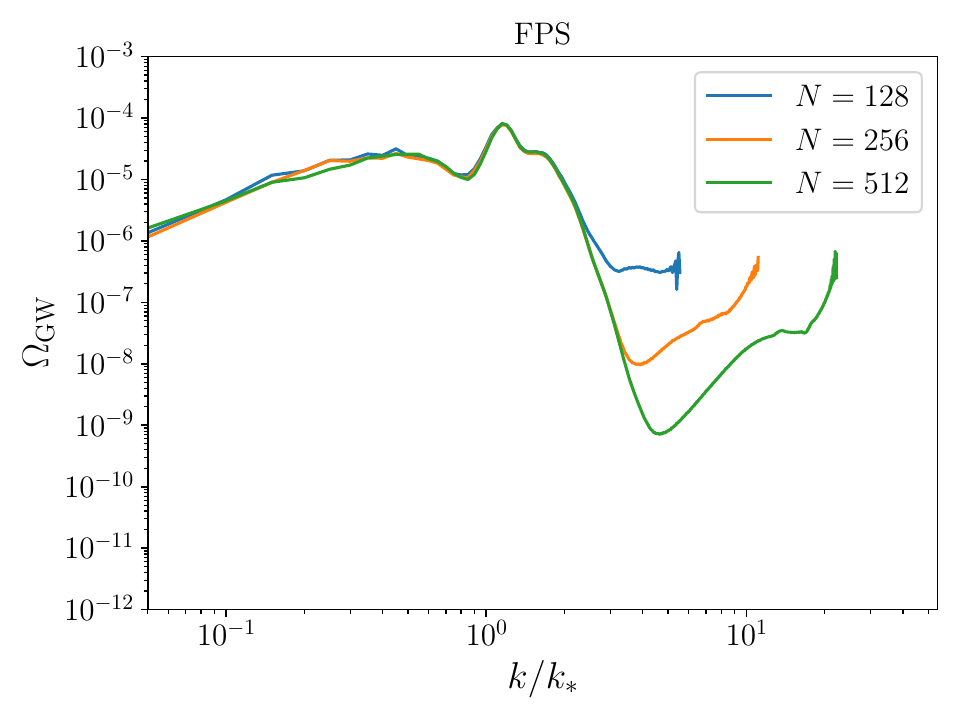}
    \includegraphics[width=0.32\linewidth]{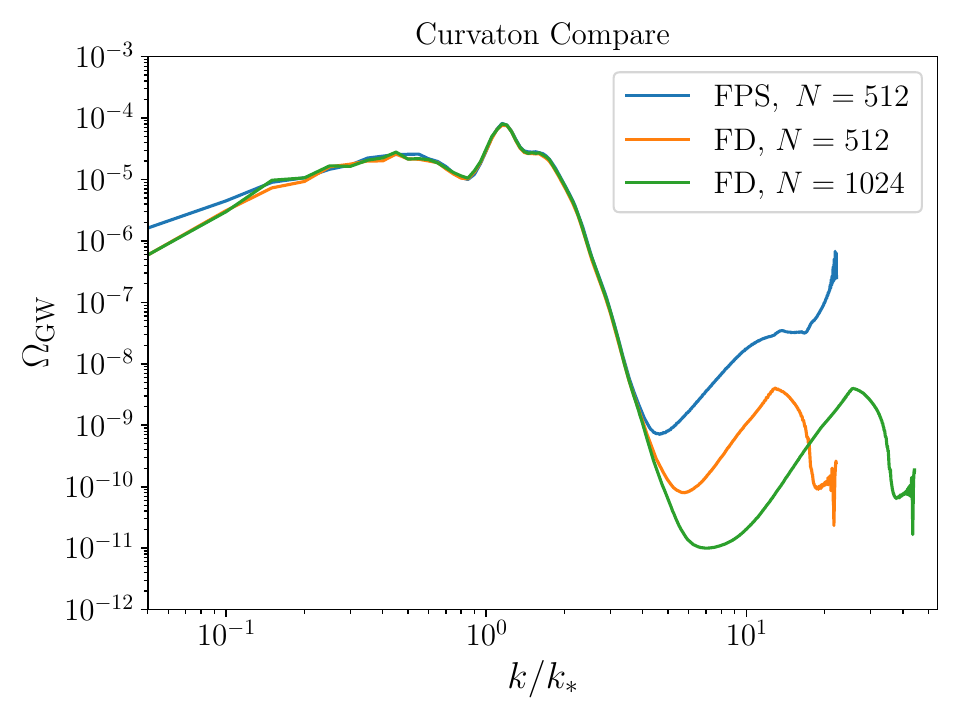}\\
    \includegraphics[width=0.32\linewidth]{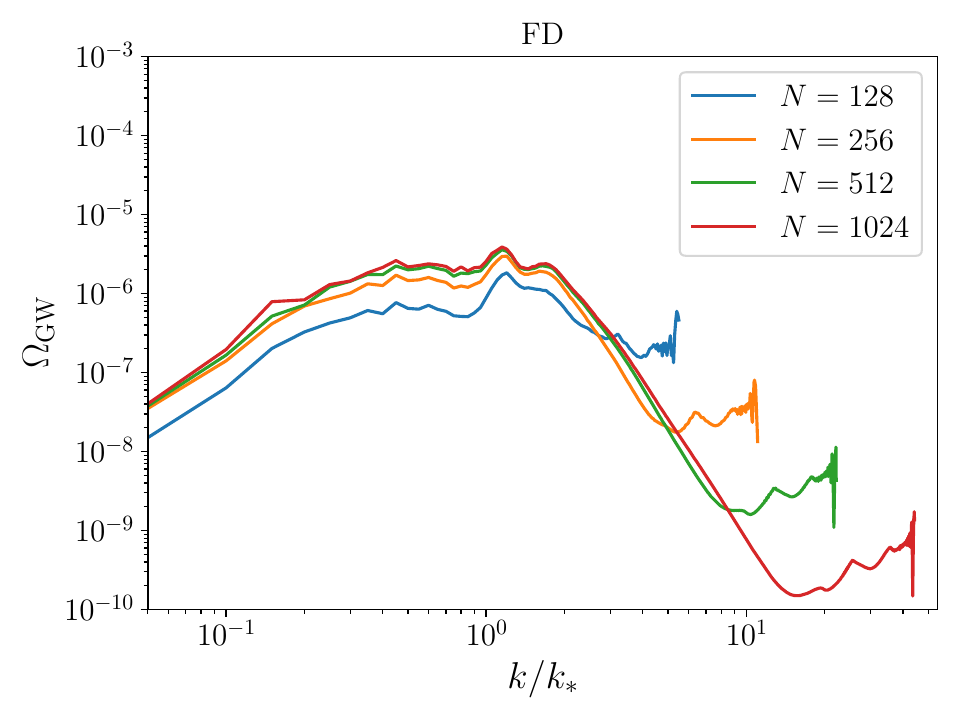}
    \includegraphics[width=0.32\linewidth]{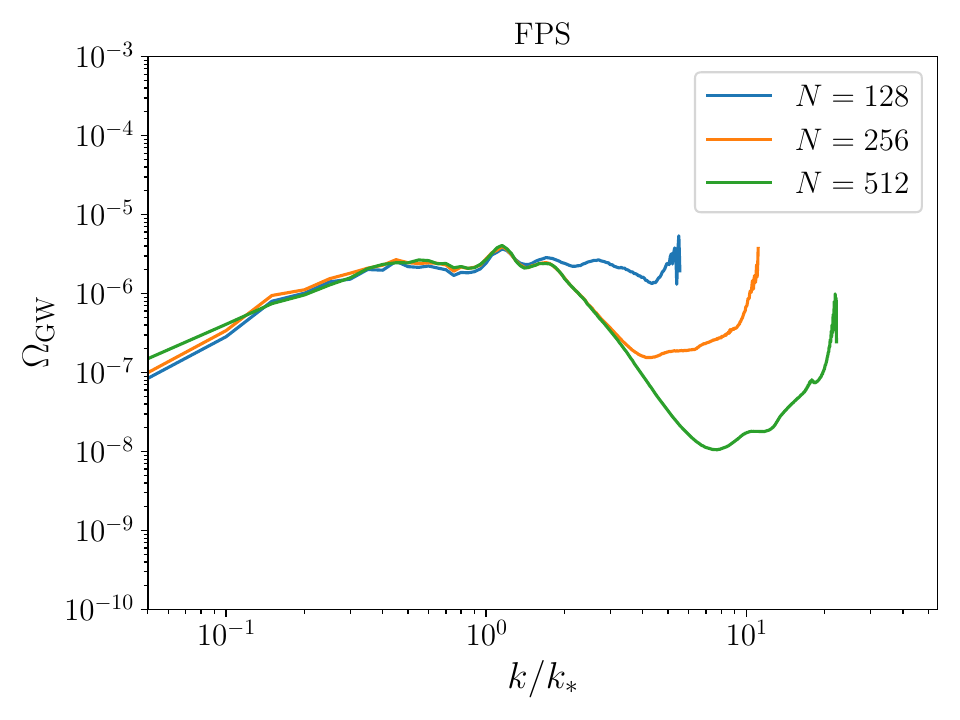}
    \includegraphics[width=0.32\linewidth]{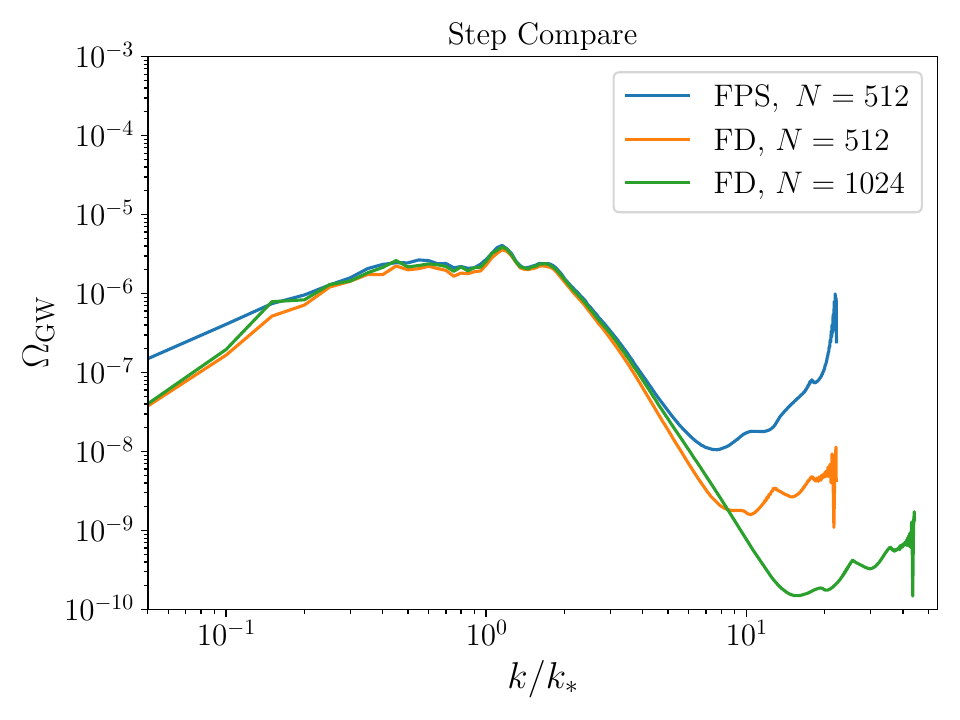}
    \caption{\black{Energy spectra of SIGW for different cases with characteristic scale $k_{*}=20$. The first, second, and third rows present the results for the logarithmic, curvaton, and step cases, respectively. The first and second columns show the results obtained using the finite difference method and the Fourier pseudospectral method, respectively, while the third column compares the results from these two methods.}}
    \label{fig:compare}
\end{figure}



\black{For the non-Gaussian case, due to the absence of relevant semi-analytical full results, we need to increase the lattice number $N^3$ to verify the convergence of the GW spectra. The primary motivation for employing the FPS method is that the dominant source of numerical error in our simulations arises from the computation of spatial derivatives. To assess the impact of spatial derivatives, we compare simulation results obtained using the FPS method with those from the FD method in the following analysis. Although implementing the FPS method in parallel is technically challenging, we adopt a hybrid approach: scalar perturbations are evolved using the fourth-order Runge-Kutta method, while tensor perturbations are evolved using the leapfrog method to facilitate simulations with $N = 512$ under our available GPU memory constraints. We have verified that both evolution schemes yield consistent results for tensor perturbations in our setup, and the leapfrog method helps alleviate memory limitations for larger resolutions. We have tested the examples of Fig. 6 of the main text.}

\begin{figure}[htbp!]
    \centering
    \includegraphics[width=0.8\linewidth]{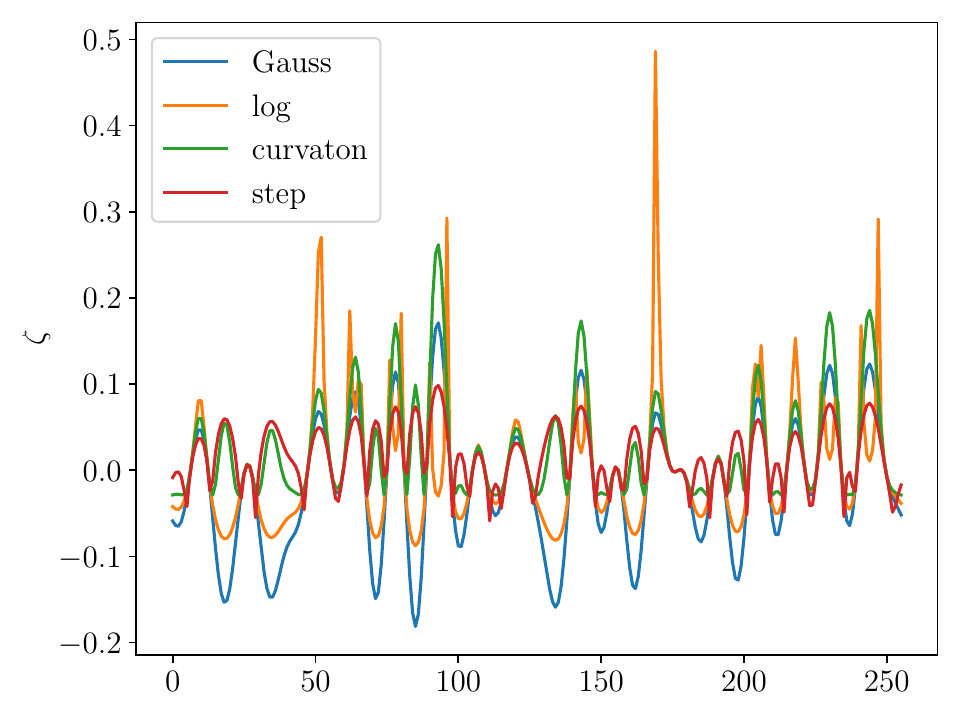}
    \caption{\black{The example slice of the initial value of curvature perturbation $\zeta$ with $k_{*}=20, N=256$.}}
    \label{fig:slice}
\end{figure}

\black{As shown in Fig.~\ref{fig:compare}, except for the logarithmic case, convergent results can be achieved with $N=256$ using the FPS method, whereas the FD method requires $N=512$ to reach convergence. Fig.~\ref{fig:slice} displays an example slice of the initial curvature perturbations. In the logarithmic case, the presence of a divergent pole and an absolute value leads to very sharp peaks in the initial curvature perturbation; consequently, obtaining accurate spatial derivatives is particularly challenging, making convergence difficult for this case. Nevertheless, we will demonstrate in the next section that by combining multiple simulation results, a convergent energy spectrum can still be obtained with a lattice number as small as $N=256$. For other well-behaved cases, such as the curvaton model, a lattice number of $N=128$ with the FPS method or $N=256$ with the FD method is sufficient to achieve a convergent energy spectrum.}

\section{Appendix C. \black{The simulation failure and combination strategy}}

\begin{figure}[h]
    \centering
    \includegraphics[width=0.32\linewidth]{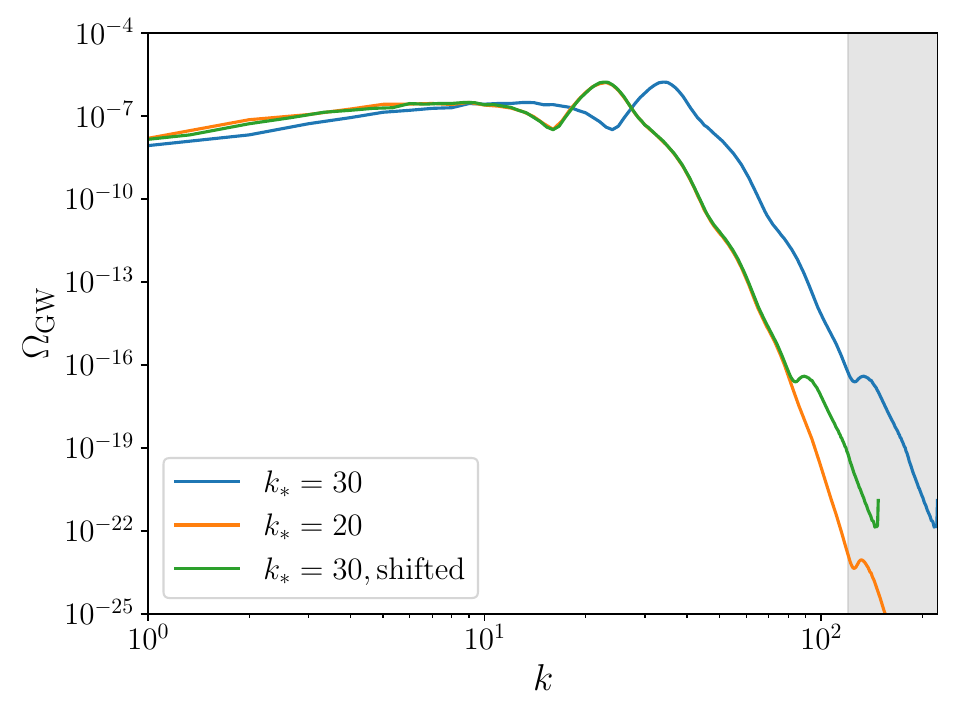}
    \includegraphics[width=0.32\linewidth]{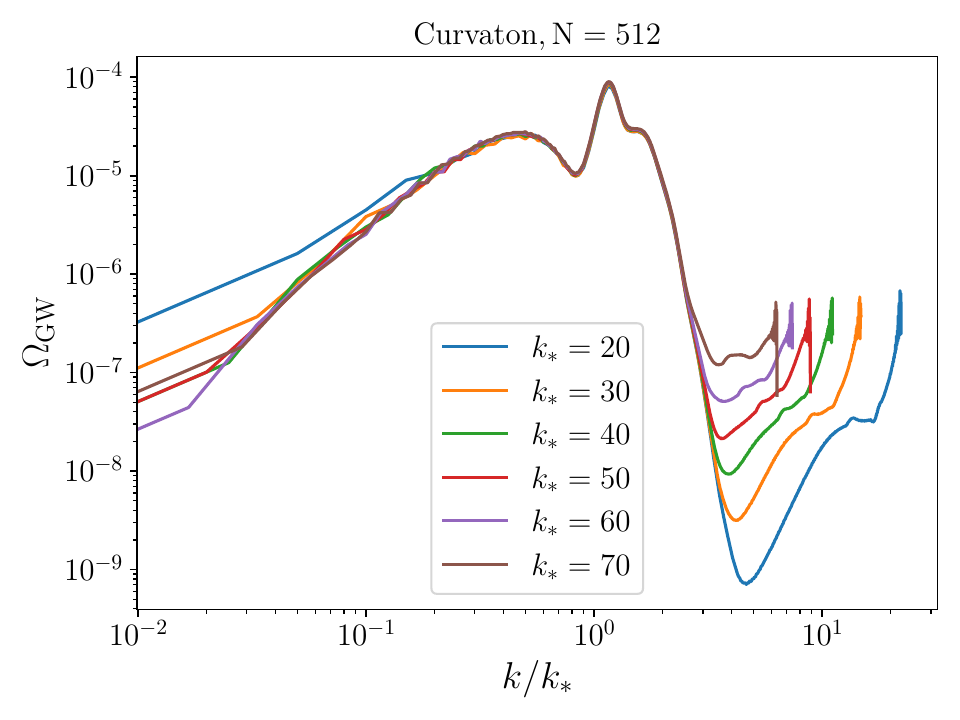}
    \includegraphics[width=0.32\linewidth]{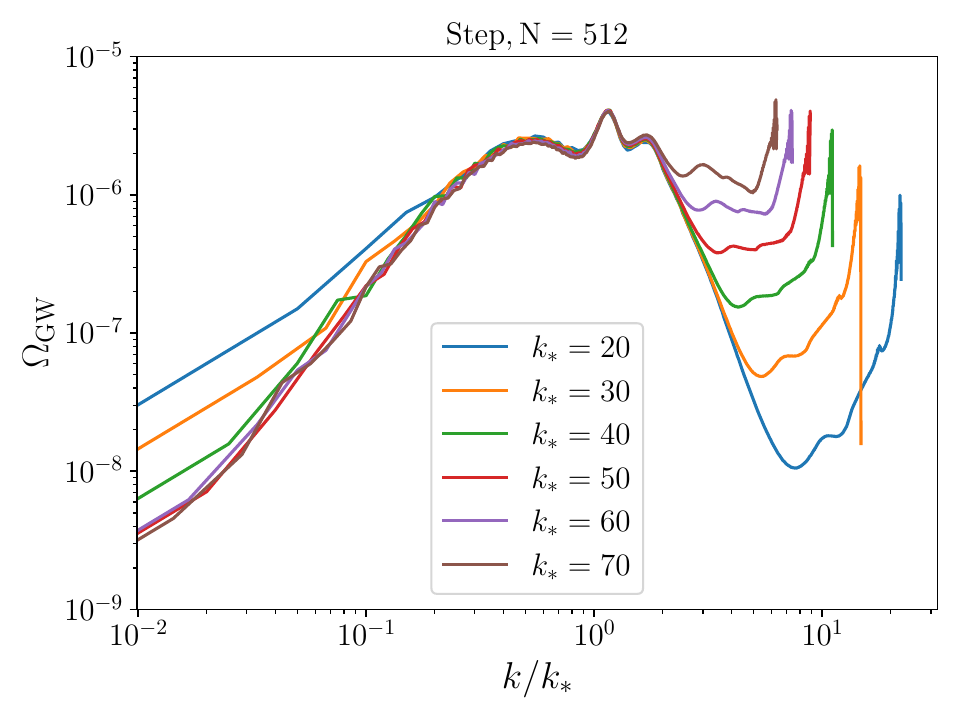}
    \caption{\black{Energy spectra of SIGW for different characteristic scale $k_{*}$ and models by using FPS method. \textit{Left}: Simulations are done with the lattice number $N^3=256^3$. The green curve means that we have shifted the $k_{*} =30$ case to match the $k_{*}=20$ case. \textit{Middle}: Curvaton model with the same parameters in Appendix~\ref{app:B}B. \textit{Right}: Step model with the same parameters in Appendix~\ref{app:B}B.}}
    \label{fig:error}
\end{figure}

\black{In our lattice simulations, the fields are evolved within a comoving box with a fixed characteristic size scale $k_*$. By normalizing the largest scale in the box to $k = 1$ and $\mathrm{d}k=1$, varying $k_*$ effectively adjusts the resolution across different scales: a larger value of $k_*$ allocates more grid points to the infrared region. Consequently, there are two approaches to improve the resolution in the ultraviolet regime: (i) increasing the lattice number $N^3$, or (ii) choosing a smaller characteristic scale. These two strategies can also be employed to identify the scale at which the simulation becomes unreliable: by comparing results obtained with different characteristic scales or lattice resolutions, one can determine the scale beyond which these different simulation results diverse. For example, as shown in the left picture of Fig.~\ref{fig:error}, the mismatch between the green and orange curves is the scale of simulation failure.} 

\black{Moreover, combining simulation results obtained with different characteristic scales allows for a more accurate reconstruction of the gravitational wave energy spectrum over a wider range of scales. As shown in the middle and right panels of Fig.~\ref{fig:error}, a smaller $k_*$ has better convergence in the UV than in the IR regimes, while a larger $k_*$ has better convergence in the IR than in the UV regimes. This suggests that a combination of two simulations of different characteristic scales would render a convergent result for the whole length regime.}


\begin{figure}[h]
    \centering
    \includegraphics[width=0.32\linewidth]{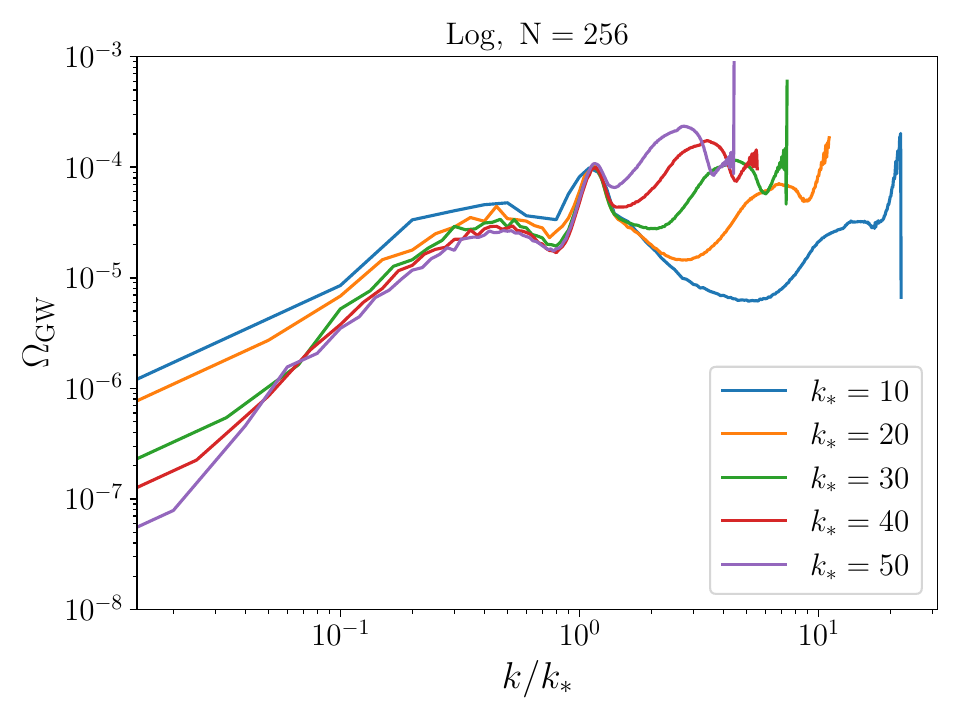}
    \includegraphics[width=0.32\linewidth]{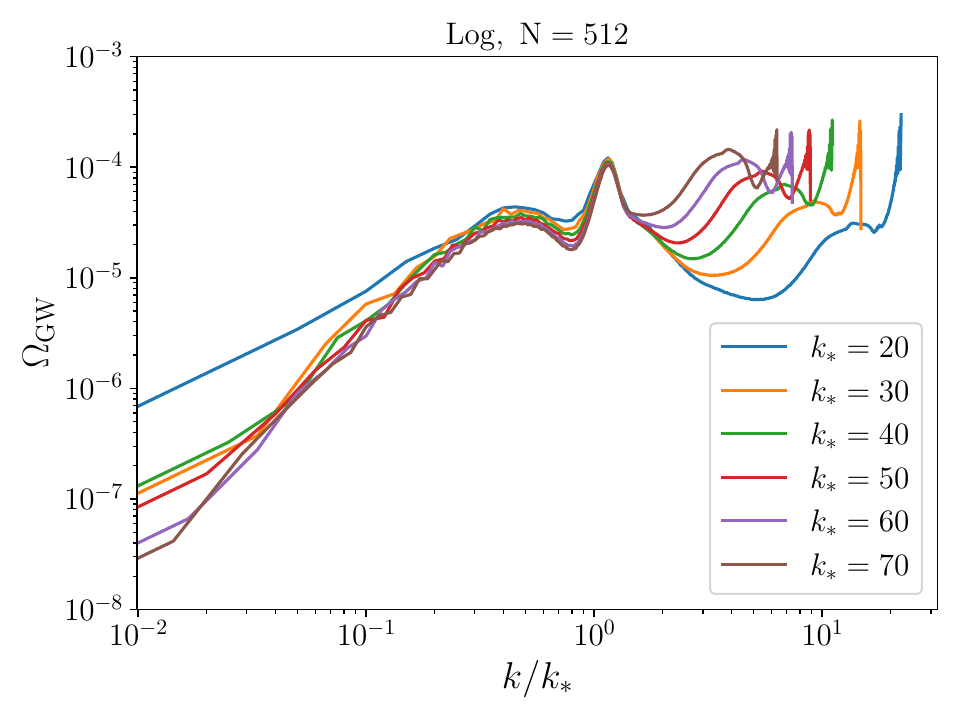}
    \includegraphics[width=0.32\linewidth]{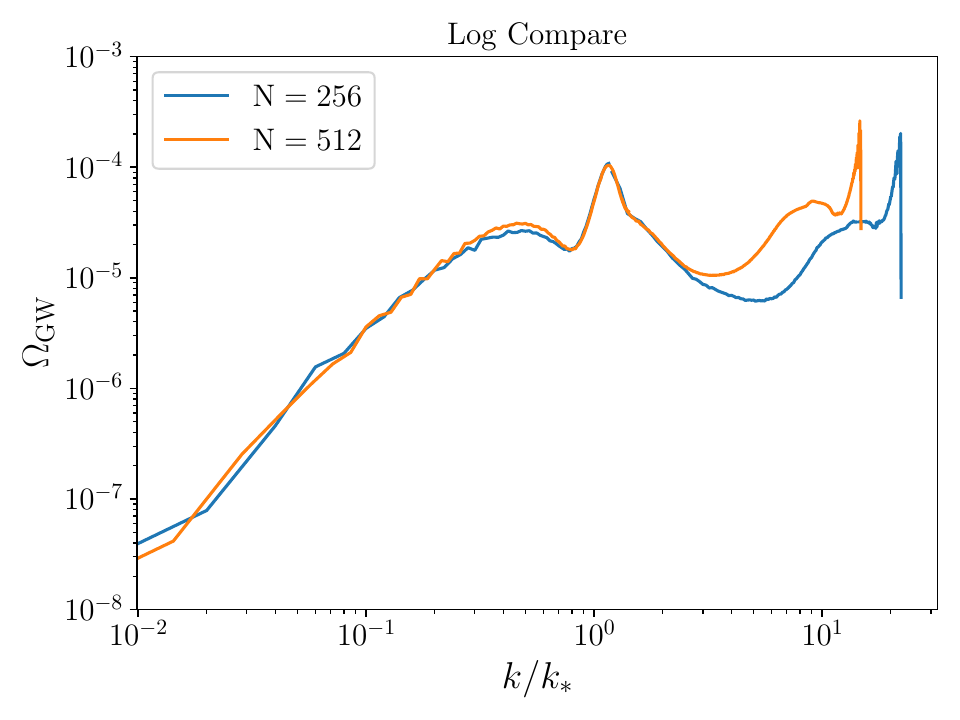}
    \caption{\black{Energy spectra of SIGW for the logarithmic case using the FPS method, shown for different characteristic scales $k_{*}$ and lattice numbers. The simulation parameters are the same as those in Appendix~\ref{app:B}B. \textit{Left}: Simulations with lattice number $N^3 = 256^3$. \textit{Middle}: Simulations with lattice number $N^3 = 512^3$. \textit{Right}: For $N = 256$, the energy spectrum is obtained by combining the results for $k_* = 10$ and $k_* = 50$; for $N = 512$, the spectrum is obtained by combining the results for $k_* = 30$ and $k_* = 70$.}}
    \label{fig:logerror}
\end{figure}

\black{In Fig.~\ref{fig:logerror}, we have shown that one can obtain the convergent energy spectrum for the logarithmic case by combining several simulation results of different characteristic scales $k_*$, even with a lattice number as small as $N^3=256^3$ with the FPS method, compared to the combination result with $N^3=512^3$.}

\section{Appendix D. Curvaton model}
For the curvaton model, we will follow the notation of Refs.~\cite{Sasaki:2006kq, Pi:2021dft, Ferrante:2022mui}. The fully nonlinear curvature perturbation is expressed as
\begin{align}\label{eq:curvaton}
    \zeta = \log [X(r, \zeta_{g})]
\end{align}
with
\begin{align}
    X(r, \zeta_g) = K^{1/2}\frac{1 + \sqrt{ArK^{-3/2} - 1}}{(3+r)^{1/3}},
\end{align}
\begin{align}
    K = \frac{1}{2}\left( P^{1/3} + (r-1)(3+r)^{1/3}P^{-1/3}  \right),
\end{align}
\begin{align}
    P = (Ar)^2 + \sqrt{(Ar)^4 + (3+r)(1-r)^3},
\end{align}
\begin{align}
    A = \frac{(2r + 3\zeta_g)^2}{4r^2},
\end{align}
where the energy fraction of curvaton to the total energy density at the time of curvaton decay, $r$, is defined by
\begin{align}
    r = \frac{3\rho_{\chi}}{3\rho_{\chi} + 4\rho_{\gamma}}.
\end{align}
Therefore, one can easily find
\begin{align}
    \log[X(1, \zeta_g)] = \frac{2}{3}\log\left( 1 + \frac{3}{2}\zeta_g\right).
\end{align}
By expanding the formula~\eqref{eq:curvaton} in terms of $\zeta_g$, one arrives at
\begin{align}
    F_{\mathrm{NL}} =  \frac{3}{4r} -1 -\frac{r}{2} .
\end{align}

\section{\black{Appendix E. Discussion on the approximate UV power law behavior}}
\begin{figure}[h]
    \centering
    \includegraphics[width=0.48\linewidth]{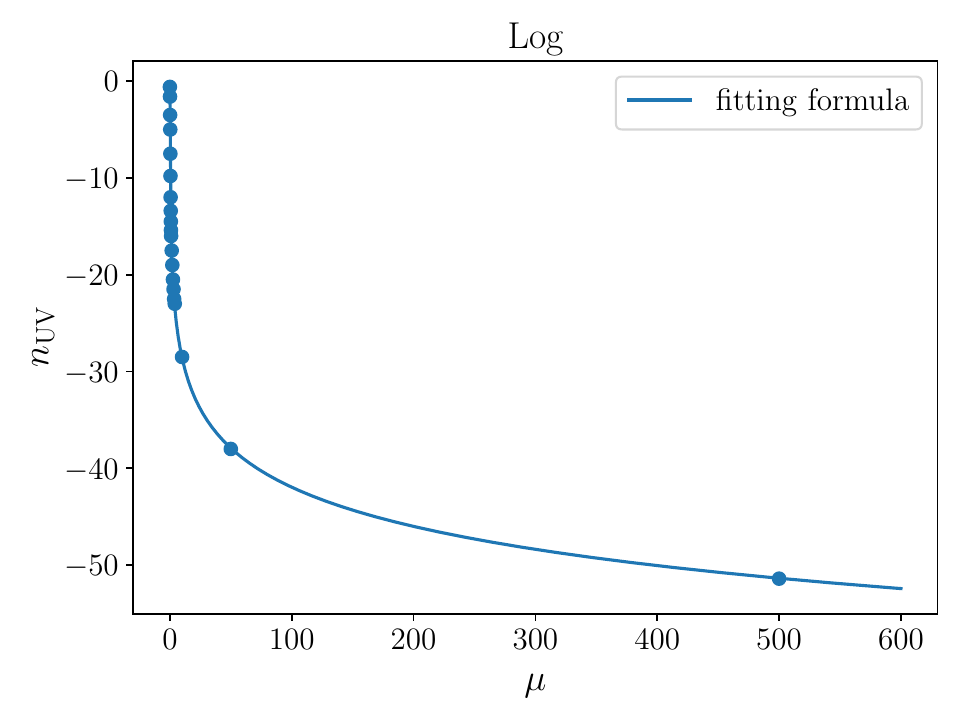}
    \includegraphics[width=0.48\linewidth]{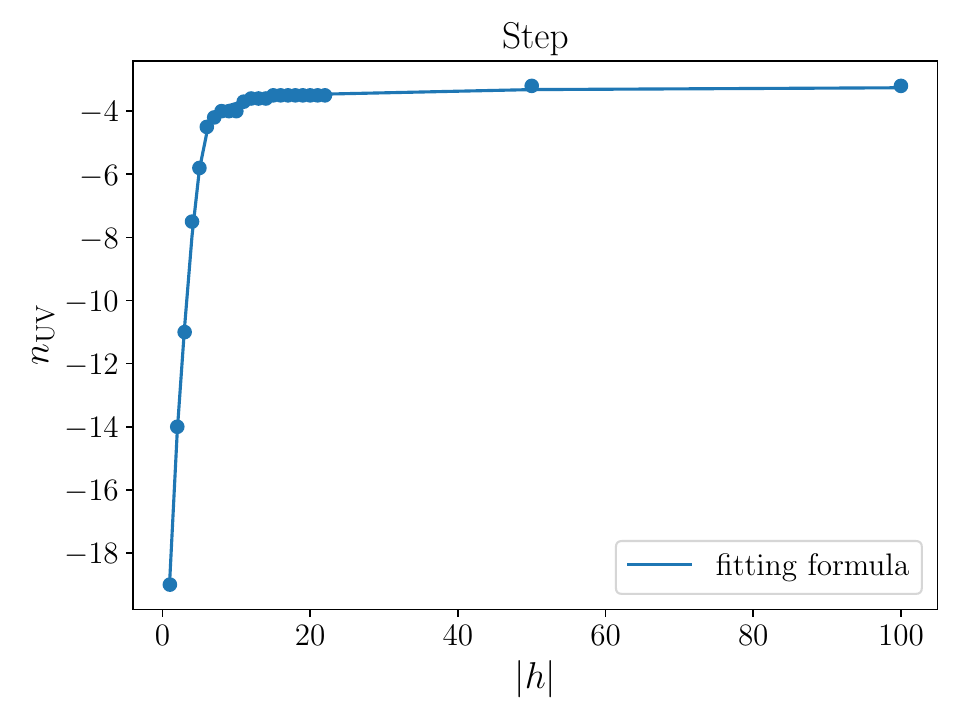}\\
    \includegraphics[width=0.48\linewidth]{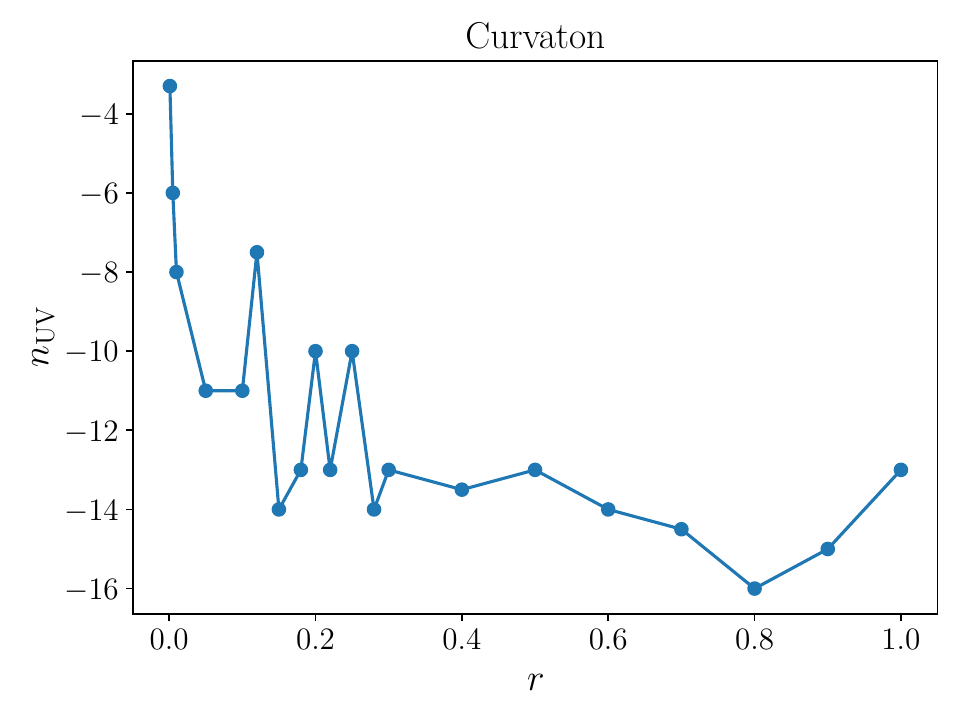}
    \includegraphics[width=0.48\linewidth]{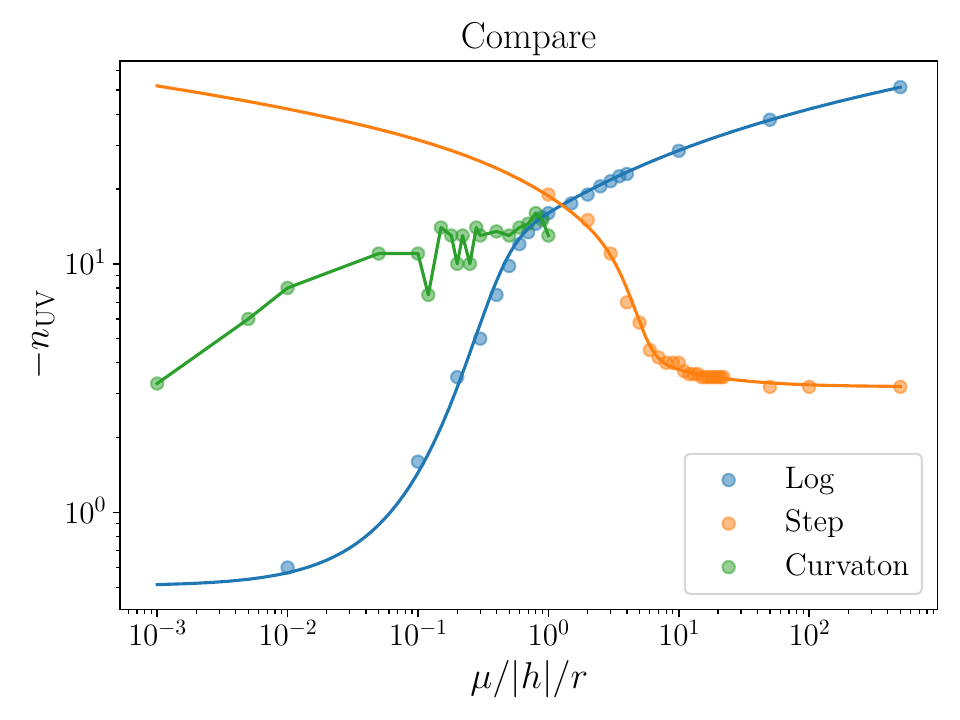}
    \caption{\black{The relation between the approximate UV power law and the non-Gaussian parameters. The scatter points are data from the simulation, while the curves for the logarithmic model and step model are the fitting lines.}}
    \label{fig:fit}
\end{figure}

\black{As discussed in the main text, the distinct UV behaviors of different models can serve as a tool to distinguish among various inflation models. Therefore, providing a fitting formula for the approximate UV power law would be highly valuable. However, this is a nontrivial task, as the parameter space is vast: the UV behavior is sensitive not only to the amplitude and shape of the power spectrum but also to the non-Gaussian parameter of the specific model. As a preliminary attempt, we fix the amplitude of the Gaussian bump power spectrum to $A = 1$ and its width to $e = 0.1$, and provide fitting formulas for the logarithmic and step models. Although we are unable to obtain a fitting formula for the curvaton model, we present a scatter plot to facilitate comparison with the other models, as shown in Fig.~\ref{fig:fit}.}
\black{The fitting formula for the logarithmic model is
\begin{align}
    n_{\mathrm{UV}} = - (3.9\ln{(\mu + 0.2)} + 10)\frac{3}{2 + e^{3-8\mu}},
\end{align}
while the step model is
\begin{align}
    n_{\mathrm{UV}} = -3.2 - \frac{6}{|h|+1} + (4\ln|h| - 12)\frac{1.1}{1 + e^{x-4}}.
\end{align}
}
\black{We emphasize that the UV power law is only approximate; consequently, its exponent may vary by as much as $\pm 1$ depending on the fitting choices. This suggests that deriving reliable fitting formulas may not be a practical approach. Fortunately, we find that a convergent GW energy spectrum can be obtained with a lattice number $N^3 = 256^3$ using the FPS method, which makes the simulation feasible on a personal computer. Therefore, performing the simulation directly may be a more reliable alternative.}


\end{document}